%% file: main.tex
\documentclass[sigconf,nonacm]{acmart}

\AtBeginDocument{%
  }

\setcopyright{acmlicensed}
\copyrightyear{2025}
\acmYear{2025}
\acmDOI{XXXXXXX.XXXXXXX}
\acmConference[Conference acronym 'XX]{Make sure to enter the correct
  conference title from your rights confirmation email}{June 03--05,
  2025}{Woodstock, NY}
\acmISBN{978-1-4503-XXXX-X/202/06}

\input{preamble}

\input{datalog-results}

\newcommand{\toolName}{XChainWatcher}

\begin{document}

\title{\toolName: Identifying Anomalies in Cross-Chain Bridges}

\settopmatter{authorsperrow=3}

\author{André Augusto}
\affiliation{%
  \institution{INESC-ID \& IST, University of Lisbon}
  \city{Lisbon}
  \country{Portugal}
}

\author{Rafael Belchior}
\affiliation{%
  \institution{INESC-ID \& Blockdaemon}
  \city{Dublin}
  \country{Ireland}
}

\author{Jonas Pfannschmidt}
\affiliation{%
  \institution{Blockdaemon}
  \city{Dublin}
  \country{Ireland}
}

\author{André Vasconcelos}
\affiliation{%
  \institution{INESC-ID \& IST, University of Lisbon}
  \city{Lisbon}
  \country{Portugal}
}

\author{Miguel Correia}
\affiliation{%
  \institution{INESC-ID \& IST, University of Lisbon}
  \city{Lisbon}
  \country{Portugal}
}

\renewcommand{\shortauthors}{Trovato et al.}

\input{sections/abstract}




\maketitle

\input{sections/introduction}

\input{sections/background}

\input{sections/solution}

\input{sections/evaluation}

\input{sections/limitations-and-future-work}

\input{sections/related-work}

\input{sections/conclusion}


\bibliographystyle{ACM-Reference-Format}
\bibliography{main}

\appendix

\end{document}

%% file: preamble.tex
\usepackage{xstring}
\usepackage{adjustbox}
\usepackage{amsfonts}
\usepackage{booktabs}
\usepackage{hhline}
\usepackage{color}
\usepackage[switch]{lineno}
\usepackage{mathtools}
\usepackage{pifont}
\usepackage{multirow}
\usepackage{colortbl}
\usepackage{tabularx}
\usepackage{arydshln}
\usepackage{enumitem,kantlipsum}
\usepackage[dvipsnames,table]{xcolor}
\usepackage{tikz}
\usepackage{comment}
\usepackage[normalem]{ulem}
\usepackage{mdwlist}
\useunder{\uline}{\ul}{}
\usepackage{threeparttable}

\usepackage{hyperref}
\usepackage{xurl}
\usetikzlibrary{arrows, positioning, automata, fit}
\usepackage{pgf}
\usepackage[most]{tcolorbox}
\usepackage{fancyvrb}

\newif\ifcomments
\commentstrue

\newcommand{\featheader}[1]{\emph{\textbf{{#1}.}}}

\definecolor{asparagus}{rgb}{0.53, 0.8, 0.42}

\definecolor{greenListing}{rgb}{0.0, 0.5, 0.0}

\definecolor{bleudefrance}{rgb}{0.19, 0.55, 0.91}

\definecolor{cadmiumorange}{rgb}{0.93, 0.53, 0.18}

\definecolor{bittersweet}{rgb}{1.0, 0.44, 0.37}

\definecolor{bostonuniversityred}{rgb}{0.65, 0.0, 0.0}

\definecolor{red1}{rgb}{167,0,0}
\definecolor{red2}{rgb}{255,0,0}
\definecolor{red3}{rgb}{255,82,82}
\definecolor{red4}{rgb}{255,123,123}
\definecolor{red5}{rgb}{255,186,186}

\usepackage{listings}

\lstdefinelanguage{Solidity}{
  keywords={contract, function, returns, TokenDeposited},
  keywordstyle=\color{cadmiumorange}\bfseries,
  ndkeywords={address, require, uint256, uint64, bool, event},
  ndkeywordstyle=\color{bleudefrance}\bfseries,
  identifierstyle=\color{black},
  sensitive=false,
  comment=[l]{//},
  morecomment=[s]{/*}{*/},
  commentstyle=\color{greenListing}\ttfamily,
  stringstyle=\color{black}\ttfamily,
  morestring=[b]',
  morestring=[b]",
}

\lstdefinestyle{solidity}{
    backgroundcolor=\color{white},
    commentstyle=\color{greenListing},
    keywordstyle=\color{blue},
    numberstyle=\color{black},
    stringstyle=\color{black},
    basicstyle=\footnotesize\ttfamily,
    breakatwhitespace=false,
    breaklines=true,
    captionpos=b,
    keepspaces=true,
    showspaces=false,
    showstringspaces=false,
    showtabs=false,
    keepspaces=true,
    frame=tb, 
    escapeinside={(*@}{@*)}, 
}

\definecolor{PrologPredicate}{RGB}{000,031,255}

\lstdefinelanguage{Datalog}{
  comment=[l]{//},
  keywords={.decl, Transaction, LockAsset, MintAsset, ValidCCTX_Rule1, native_deposit, native_withdrawal, sc_token_deposited, tc_token_deposited, tc_token_withdrew, sc_token_withdrew, erc20_transfer, transaction, bridge_controlled_address, token_mapping, wrapped_native_token, TC_ValidERC20TokenDeposit, SC_ValidERC20TokenDeposit, SC_ValidNativeTokenDeposit, CCTX_ValidDeposit, CCTX_ValidWithdrawal, TC_ValidERC20TokenWithdrawal, TC_ValidNativeTokenWithdrawal, SC_ValidERC20TokenWithdrawal, sc_withdrawal, cctx_finality},
  keywordstyle=\color{PrologPredicate},
  identifierstyle=\color{black},
  sensitive=false,
}

\lstdefinestyle{datalog}{
    backgroundcolor=\color{white},
    commentstyle=\color{greenListing},
    keywordstyle=[1]\color{blue},
    keywordstyle=[2]\color{cadmiumorange},
    stringstyle=\color{black},
    basicstyle=\scriptsize\ttfamily,
    breakatwhitespace=false,
    breaklines=true,
    captionpos=b,
    keepspaces=true,
    showspaces=false,
    showstringspaces=false,
    showtabs=false,
    keepspaces=true,
    frame=tb, 
    escapeinside={(*@}{@*)}, 
}

\newcommand{\truncate}[1]{%
  \StrLeft{#1}{6}...\StrRight{#1}{4}
}

\newcommand{\AddrHref}[3][blue]{\href{#2}{\color{#1}{\truncate{#3}}}}%

\newcommand{\Keyword}[1]{\texttt{{\small#1}}}

\newcommand{\FootnotesizeKeyword}[1]{\texttt{{\small#1}}}
\newcommand{\ScriptsizeKeyword}[1]{\texttt{{\scriptsize#1}}}

\newcommand{\RuleName}[1]{\texttt{{\small\color{PrologPredicate}{#1}}}}
\newcommand{\FootNoteSizeRuleName}[1]{\texttt{{\footnotesize\color{PrologPredicate}{#1}}}}
\newcommand{\scriptSizeRuleName}[1]{\texttt{{\scriptsize\color{PrologPredicate}{#1}}}}

\newcommand\CCTX{$\mathit{cctx}$}
\newcommand\CCTXS{$\mathit{cctxs}$}

\definecolor{networks}{RGB}{144, 194, 104}
\definecolor{networks_outline}{HTML}{2D7600}

\newcommand*\circled[2]{\tikz[baseline=(char.base)]{
\node[circle,draw=#2,scale=0.8,inner sep=2pt] (char) {\color{#2}#1};}}

\newtcolorbox{anomalybox}[1][]{%
  colback=gray!5,      
  colframe=black,    
  boxrule=1pt,       
  arc=0pt,             
  auto outer arc,
  left=2mm, right=2mm,
  top=1mm, bottom=1mm,
  fontupper=\footnotesize,
  #1
}

%% file: datalog-results.tex
\newcommand{\SCDepositFactsNomad}{7,187~}

\newcommand{\ERCTransferSDepositsFactsNomad}{4,263~}
\newcommand{\SCTokenDepositedFactsNomad}{11,411~}

\newcommand{\SCValidNativeTokenDepositRoninCaptured}{38,462~}
\newcommand{\SCValidNativeTokenDepositRoninUnmatched}{10~}
\newcommand{\SCValidERCTokenDepositRoninCaptured}{5,527~}
\newcommand{\SCValidERCTokenDepositRoninUnmatched}{0~}
\newcommand{\TCValidERCTokenDepositRoninCaptured}{43,990~}

\newcommand{\TCValidERCTokenWithdrawalRoninCaptured}{35,411~}

\newcommand{\TCValidNativeTokenWithdrawalRoninCaptured}{0~}
\newcommand{\TCValidNativeTokenWithdrawalRoninUnmatched}{0~}
\newcommand{\SCValidERCTokenWithdrawalRoninCaptured}{25,470~}

\newcommand{\CCTXDepositNomadCaptured}{11,404~}

\newcommand{\SCValidNativeTokenDepositNomadCaptured}{7,187~}
\newcommand{\SCValidNativeTokenDepositNomadUnmatched}{0~}
\newcommand{\SCValidERCTokenDepositNomadCaptured}{4,223~}
\newcommand{\SCValidERCTokenDepositNomadUnmatched}{6~}
\newcommand{\TCValidERCTokenDepositNomadCaptured}{11,417~}
\newcommand{\TCValidERCTokenDepositNomadUnmatched}{13~}

\newcommand{\TCValidERCTokenWithdrawalNomadCaptured}{4,846~}

\newcommand{\TCValidNativeTokenWithdrawalNomadCaptured}{464~}

\newcommand{\SCValidERCTokenWithdrawalNomadCaptured}{4,869~}
\newcommand{\SCValidERCTokenWithdrawalNomadUnmatched}{387~}
\newcommand{\TCValidNativeTokenWithdrawalAddDataNomadUnmatched}{238}
\newcommand{\TCValidERCTokenWithdrawalAddDataNomadUnmatched}{491}

\newcommand{\totaldstaddressesinwithdrawalswithnomatchNomad}{729~}
\newcommand{\totaldstaddressesinwithdrawalswithnomatchnobalanceNomad}{121~}
\newcommand{\totaldstaddressesinwithdrawalswithnomatchunderfeeNomad}{231~}
\newcommand{\totaldstaddressesinwithdrawalswithnomatchnobalanceandunderfeeNomad}{72~}
\newcommand{\totaldstaddressesinwithdrawalswithnomatchtotalvalueNomad}{3.62M~}
\newcommand{\totaldstaddressesinwithdrawalswithnomatchsingletriesNomad}{592~}
\newcommand{\totaldstaddressesinwithdrawalswithnomatchmultipletriesNomad}{58~}

\newcommand{\totaldstaddressesinwithdrawalswithnomatchbeforeNomad}{541~}
\newcommand{\totaldstaddressesinwithdrawalswithnomatchbeforenobalanceNomad}{95~}
\newcommand{\totaldstaddressesinwithdrawalswithnomatchbeforeunderfeeNomad}{185~}
\newcommand{\totaldstaddressesinwithdrawalswithnomatchbeforenobalanceandunderfeeNomad}{55~}
\newcommand{\totaldstaddressesinwithdrawalswithnomatchbeforetotalvalueNomad}{0.34M~}
\newcommand{\totaldstaddressesinwithdrawalswithnomatchbeforesingletriesNomad}{460~}
\newcommand{\totaldstaddressesinwithdrawalswithnomatchbeforemultipletriesNomad}{34~}
\newcommand{\totaldstaddressesinwithdrawalswithnomatchafterNomad}{188~}
\newcommand{\totaldstaddressesinwithdrawalswithnomatchafternobalanceNomad}{26~}
\newcommand{\totaldstaddressesinwithdrawalswithnomatchafterunderfeeNomad}{46~}
\newcommand{\totaldstaddressesinwithdrawalswithnomatchafternobalanceandunderfeeNomad}{17~}
\newcommand{\totaldstaddressesinwithdrawalswithnomatchaftertotalvalueNomad}{3.27M~}
\newcommand{\totaldstaddressesinwithdrawalswithnomatchaftersingletriesNomad}{136~}
\newcommand{\totaldstaddressesinwithdrawalswithnomatchaftermultipletriesNomad}{23~}

\newcommand{\totaldstaddressesinwithdrawalswithnomatchRonin}{11,794~}
\newcommand{\totaldstaddressesinwithdrawalswithnomatchnobalanceRonin}{6,054~}
\newcommand{\totaldstaddressesinwithdrawalswithnomatchunderfeeRonin}{7,469~}
\newcommand{\totaldstaddressesinwithdrawalswithnomatchnobalanceandunderfeeRonin}{5,261~}
\newcommand{\totaldstaddressesinwithdrawalswithnomatchtotalvalueRonin}{1.18M~}
\newcommand{\totaldstaddressesinwithdrawalswithnomatchsingletriesRonin}{9,657~}
\newcommand{\totaldstaddressesinwithdrawalswithnomatchmultipletriesRonin}{956~}

\newcommand{\totaldstaddressesinwithdrawalswithnomatchbeforeRonin}{11,574~}
\newcommand{\totaldstaddressesinwithdrawalswithnomatchbeforenobalanceRonin}{5,988~}
\newcommand{\totaldstaddressesinwithdrawalswithnomatchbeforeunderfeeRonin}{7,381~}
\newcommand{\totaldstaddressesinwithdrawalswithnomatchbeforenobalanceandunderfeeRonin}{5,212~}
\newcommand{\totaldstaddressesinwithdrawalswithnomatchbeforetotalvalueRonin}{1.09M~}
\newcommand{\totaldstaddressesinwithdrawalswithnomatchbeforesingletriesRonin}{9,490~}
\newcommand{\totaldstaddressesinwithdrawalswithnomatchbeforemultipletriesRonin}{932~}
\newcommand{\totaldstaddressesinwithdrawalswithnomatchafterRonin}{220~}
\newcommand{\totaldstaddressesinwithdrawalswithnomatchafternobalanceRonin}{66~}
\newcommand{\totaldstaddressesinwithdrawalswithnomatchafterunderfeeRonin}{88~}
\newcommand{\totaldstaddressesinwithdrawalswithnomatchafternobalanceandunderfeeRonin}{49~}
\newcommand{\totaldstaddressesinwithdrawalswithnomatchaftertotalvalueRonin}{0.09M~}
\newcommand{\totaldstaddressesinwithdrawalswithnomatchaftersingletriesRonin}{176~}
\newcommand{\totaldstaddressesinwithdrawalswithnomatchaftermultipletriesRonin}{21~}

%% file: sections/abstract.tex
\begin{abstract}

Cross-chain bridges are a type of middleware for blockchain interoperability that supports the transfer of assets and data across blockchains. 
However, several of these bridges have vulnerabilities that have caused 3.2 billion dollars in losses since May 2021. Some studies have revealed the existence of these vulnerabilities, but there is little quantitative research available and there are no safeguard mechanisms to protect bridges from such attacks. Furthermore, no studies are available on the practices of cross-chain bridges that can cause financial losses. 
We propose \toolName~(Cross-Chain Watcher), a modular and extensible logic-driven anomaly detector for cross-chain bridges. It operates in three main phases: (1) decoding events and transactions from multiple blockchains, (2) building logic relations from the extracted data, and (3) evaluating these relations against a set of detection rules. Using \toolName, we analyze data from two previously attacked bridges: the Ronin and Nomad bridges. \toolName~was able to successfully identify the transactions that led to losses of \$611M and \$190M (USD) and surpassed the results obtained by a reputable security firm in the latter. We not only uncover successful attacks, but also reveal other anomalies, such as 37 cross-chain transactions (\CCTX) that these bridges should not have accepted, failed attempts to exploit Nomad, over \$7.8M worth of tokens locked on one chain but never released on Ethereum, and \$200K lost by users due to inadequate interaction with bridges. We provide the first open dataset of 81,000 \CCTXS~across three blockchains, capturing more than \$4.2B in token transfers.
\end{abstract}

%% file: sections/introduction.tex
\section{Introduction}
\label{sec:intro}

In recent years, there has been a remarkable adoption of \emph{blockchain interoperability} through the use of cross-chain mechanisms \cite{augusto_sok_2024, belchior2021survey}. The most popular mechanisms are \emph{cross-chain bridges} (or simply \emph{bridges}). Bridges serve as an essential middleware in the blockchain ecosystem, connecting decentralized applications across various blockchains, and facilitating the transfer and exchange of assets.

In the Ethereum ecosystem, numerous bridges connect Ethereum to other blockchains, such as rollups and sidechains. Native bridges support rollups -- Layer 2 solutions designed to enhance Ethereum’s scalability while inheriting its security (e.g.,~\cite{optimism, Base, arbitrum, mantle}). In contrast, non-native bridges connect Ethereum to sidechains, which employ an independent consensus mechanism and do not inherit Ethereum’s security guarantees. Despite these differences, the primary goal remains the same: enabling decentralized applications on Ethereum to expand to faster and more cost-efficient blockchains.

\begin{figure}
    \centering
    \includegraphics[width=0.48\textwidth]{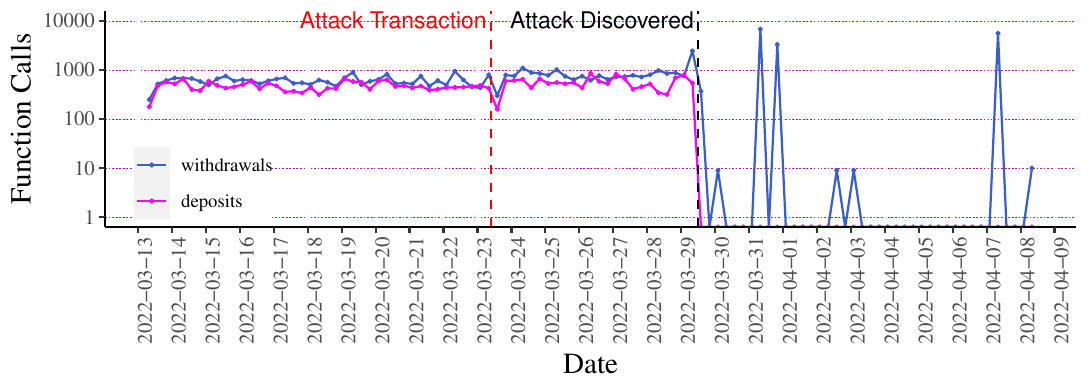}
    \caption{Number of function calls to withdraw and deposit funds into/from the Ronin blockchain through the Ronin bridge. 
    The attack was only discovered six days later, causing deposit calls to drop to zero. Each data point represents the total function calls in periods of 6 hours.}
    \label{fig: function_calls}
    \Description[<short description>]{<long description>}
\end{figure}

The cross-chain ecosystem is growing.
During 2023, cross-chain protocols raised more than \$500 million (USD) in investment rounds \cite{Blackstone_2023, Chawla, Cirrone_2023, Nijkerk_2023, Squid_2023, supra_2023}, and processed millions of cross-chain transactions (\CCTX) daily~\cite{wormhole_2023}. In November 2024, non-native cross-chain bridges had a total value locked (TVL) of around \$11 billion~\cite{l2beat} and native bridges \$39 billion, highlighting the growing interest in the technology, despite its numerous hacks: from May 2021 to August 2024, hackers stole a staggering amount of \$3.2 billion in cross-chain bridges~\cite{augusto_sok_2024}, and have indirectly caused losses of several tens of millions in other Decentralized Finance (DeFi) protocols due to on-chain activity and token valuations plummeting~\cite{nomad_rekt_2022}.


Not even extensively audited bridges are immune to vulnerabilities~\cite{wormhole_2023}. In fact, several bridges have been exploited multiple times~\cite{THORChain_hack_2021, thorchain_rekt_2021_2, thorchain_hack_2021_3, poly_network_hack_1, poly_network_hack_2, poly_network_hack_3, multichain_hack_2023, anyswap_hack_2021, ronin_chain_attack_2024, ronin_hack_rekt}. Moreover, protocols take too long to react to an attack~\cite{poly_network_hack_2, ronin_hack_rekt, pnetwork_hack}, suggesting that teams may not be sufficiently prepared to address integrity breaches, possibly due to \textbf{lack of prior awareness, observability, monitoring, or good \emph{SecOps} practices}~\cite{augusto_sok_2024}. In 2022, the Ronin Bridge was attacked, but the team discovered the attack only 6 days later (cf. Figure~\ref{fig: function_calls}). In the most recent attack, which also targeted the Ronin bridge in August 2024, the team reported that the bridge was paused only about 40 minutes after the first malicious on-chain activity was detected~\cite{ronin_chain_attack_2024}. Even if attacks cannot be reversed, it is possible to work on swift detection and protocol stoppage to avoid further exploitation (in Section~\ref{subsec: withdrawals_in_s} we show that there were 382 attacking transactions in the Nomad bridge attack). Developing effective incident response frameworks is crucial for efficient attack identification and response to malicious activity, with the great potential of minimizing losses.

Some authors have studied cross-chain security, listing and systematizing vulnerabilities and attacks across the relevant cross-chain layers~\cite{haugum_security_2022,duan2023attacks, Yin_Yan_Liang_Xie_Wan_2023,zhao_wang_yang_shang_sun_wang_yang_he_2023, zhang2023sok, augusto_sok_2024}. However, \emph{quantitative studies} with real-world data are still lacking. 
Variations in contract implementations, security models~\cite{augusto_sok_2024}, bridging models~\cite{belchior_2023_doyouneed} (e.g., \emph{lock-mint}, \emph{burn-mint}, \emph{lock-unlock}), and token types across different chains make it difficult to monitor and safeguard these systems consistently. Additionally, the use of intermediary protocols (e.g., \emph{bridge aggregators}~\cite{lifi, Subramanian_2024}) and the extraction of data from various sources (e.g., transaction data or events emitted by contracts) increase the technical challenges of performing these studies.

To address this gap, we present a middleware monitoring layer that detects anomalies in cross-chain bridges and validates them through an empirical study of real-world exploits. Many academic works suggest that anomaly detectors can be trained automatically from live-captured normal/good behavior. This approach proves impractical for cross-chain bridges because they are inherently complex systems, not formally specified, often misused, and are constantly being attacked due to the large amounts moved. Therefore, there is no hope of live capturing a clean and labeled dataset of cross-chain transactions that can be used to train anomaly detection models automatically -- and there are no open-source alternatives at the moment. In this work, to overcome this challenge, we rely on a manual definition of \textit{cross-chain rules} to characterize the expected behavior in a cross-chain bridge. These rules form the basis of our anomaly detection mechanism, enabling us to detect known and undocumented anomalies. This paper, along with the open and labeled dataset provided, establishes the first baseline for future automated approaches to cross-chain anomaly detection.

There is a large variety of bridge solutions in the industry, so designing an anomaly detection tool that fits every scenario is challenging. Therefore, in this paper, we focus on modeling and evaluating our solution for cross-chain bridges that connect Ethereum to its sidechains~\cite{sidechains}, the most valuable blockchain ecosystem except for Bitcoin (Ethereum alone has a market cap of around \$200 billion). The communication between Ethereum and a sidechain with a cross-chain bridge involves two steps: users first \textit{Deposit} assets transferring tokens from Ethereum to the sidechain, and later \textit{Withdraw} funds, transferring the assets back to Ethereum. While there are some nuances and rules that may need fine-tuning, the rationale followed in this paper can be applied to other interoperability projects (e.g., arbitrary message-passing protocols).

This paper provides the following contributions:
\begin{enumerate}
    \item \textbf{\toolName.} The first open-source framework for performing anomaly detection in cross-chain bridges, capable of detecting known attacks and other anomalies that harm users and protocol operators. \toolName~provides the pipeline for decoding event and transaction data, building logic relations, and evaluating them against the proposed anomaly detection rules.

    \item \textbf{Quantitative study on cross-chain security.} We perform an anomaly detection analysis on data extracted from bridge contracts deployed on Ethereum, Gnosis, and Moonbeam -- 3 EVM-based blockchains. 
    We release the first open dataset of cross-chain transactions, consisting of over 81,000 \CCTXS, moving more than \$$4.2$B in token transfers.

    \item \textbf{New anomalies.} Through the analysis of the anomaly detection results, we identify past attacks and also new anomalies in cross-chain bridges due to unintended behavior from users and protocols.
\end{enumerate}

The paper is structured as follows. Section~\ref{sec: background} provides background information on blockchain, smart contracts, and cross-chain bridges. Section~\ref{sec: xchainwatcher} details the design of \toolName. Sections~\ref{sec: evaluation} and~\ref{sec: results} present the experimental setup and the anomaly detection results. Section~\ref{sec: limitations_and_future_work} outlines the discussion, limitations, and future work. Finally, the related work and conclusions are given in Sections~\ref{sec: related_work} and \ref{sec: conclusion}. All monetary values presented in this paper are in US dollars.

%% file: sections/background.tex
\section{Background}
\label{sec: background}

We provide an overview of the necessary background to understand the remainder of this paper.

\subsection{Blockchain and Smart Contracts}

Consider a blockchain $B$ a sequence of blocks $\{B_1, B_2, ..., B_n\}$, where $n$ is the n\textsuperscript{th} block such that each block is cryptographically linked to the previous one. Each block contains a root of the current state trie, which holds the state of the blockchain, represented as key-value pairs $S(B_x) = \langle key, value \rangle$. Each key represents an account -- either an \textit{Externally Owned Account} (EOA) controlled by a cryptographic key pair or a \textit{Contract Account}. The latter contains code that enforces the so-called smart contracts that execute in the native virtual machine of the blockchain, e.g., the Ethereum Virtual Machine (EVM). 
The execution of $tx$ in $B_x$ changes the state $S(B_x) \overset{\textit{tx}}\rightarrow S'(B_x)$. Examples of state changes include triggering the execution of smart contracts or actions natively supported by the blockchain, such as transferring native currency or triggering internal transactions (which can recursively trigger additional state changes). Smart contracts enable the execution of code, which can define tokens -- by following common interfaces, such as the ERC-20~\cite{erc20} or ERC-721~\cite{erc721} -- or arbitrary logic, such as validating \emph{Merkle proofs} or digital signatures. Code execution may emit events that can be understood as execution logs (also called \emph{topics} in the context of transaction receipts). Events are usually representations of state changes in a certain smart contract. 

\subsection{Cross-Chain Bridge Model}\label{sec: cross-chain-bridge-model}

\begin{figure*}[ht]
    \centering
    \includegraphics[width=0.9\textwidth]{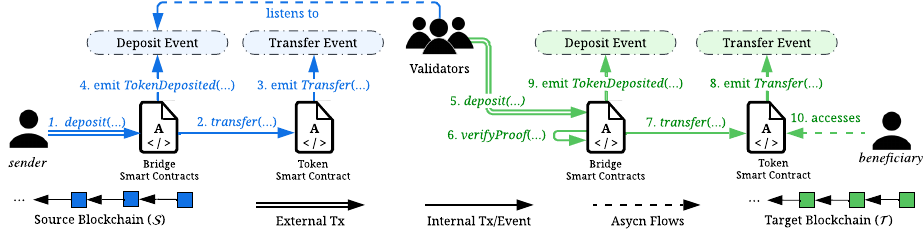}
    \caption{The flow of a token transfer from a source blockchain ($\mathcal{S}$) to a target blockchain ($\mathcal{T}$), using a cross-chain bridge.
    }
    \label{fig: cross_chain_bridge_model}
    \Description[<short description>]{<long description>}
\end{figure*}


Contrary to most DeFi dApps, bridges span over two or more blockchains, rather than being confined to one. Figure~\ref{fig: cross_chain_bridge_model} illustrates the components of a cross-chain bridge, showing a source chain (\(\mathcal{S}\)) and a target chain (\(\mathcal{T}\)) with a one-way token deposit flow (\(\mathcal{S} \rightarrow \mathcal{T}\)). 

To perform a cross-chain transfer depositing tokens into another blockchain, a user \( u_s \) issues a transaction that is added to the blockchain at timestamp \( t1 \), \( tx_{t1} \), on a source chain \( S \) to escrow tokens. This transaction can directly target a bridge contract or an intermediary protocol that calls internally a bridge contract. The bridge contract subsequently triggers an internal call to the token contract \( \tau_s \) associated with the token that \( u_s \) wishes to bridge. This call results in the creation of a commitment \( \pi_S(u_s, \tau_s, q) \), indicating that \( q \) units (quantity) of token \( \tau_s \) held by \( u_s \) have been escrowed. This commitment reflects either the locking or burning of tokens, leading to a state change in \( S \), which will be a part of the blockchain's new state \( S_{t2} \):
\[
S_{t2} \supseteq \left( S_{t1} \oplus \pi_S(u_s, \tau_s, q) \right)
\]
The state change triggers an event emission from the token contract \( \tau_s \). In this paper, we focus on the ERC20 token standard, thus, on fungible tokens~\cite{antonopoulos2018mastering}. Depending on the bridging model, escrowing tokens can be implemented by transferring tokens to a bridge-controlled address (\emph{lock} model) or to the null address (\emph{burn} model). The ability to handle this dichotomy allows this analysis to be agnostic of the bridging model. Therefore, in a \emph{lock-unlock} model, the event can be represented as:
\[
\epsilon_{\tau_s, S} = \emph{Transfer}(u_s, \emph{bridge}, q)
\]
in the form \emph{(from, to, value)}. In a \emph{burn-mint} model, tokens are instead burnt -- i.e., transferred to the null address (\emph{0x000}) -- even though not as common, this is a more secure approach because it avoids creating a honeypot of locked assets in \( \mathcal{S} \). If \( u \) is trying to bridge native tokens in $\mathcal{S}$, there is no call to the \emph{Transfer} method of an ERC20 token contract, but the commitment is in the form of a native transfer of tokens in $\mathcal{S}$ -- i.e., \( tx_{t1}.value = q\). Once the commitment is created in $\mathcal{S}$, the bridge contract emits an event with commitment data and some additional parameters, such as a unique identifier for the deposit, the beneficiary $u_t$ (the user to which the tokens are intended in $\mathcal{T}$), and the token in $\mathcal{T}$ ($\tau_t$) that represents the same token as \( \tau_s \):
\[
\epsilon_{\emph{bridge}, S} = TokenDeposited(\emph{deposit\_id}, \pi_S(u_s, \tau_s, q), u_t, \tau_t)
\]
This event is captured by validators (or relayers) -- off-chain entities responsible for enabling cross-chain interoperability. Upon detecting an event on chain \( S \), validators initiate a transaction on the target chain \( T \) to trigger a state change, such that the commitment is part of the new state \( T_{t3} \):
\[
\mathcal{T}_{t3} \supseteq \left( \mathcal{T}_{t2} \oplus \pi_{\mathcal{T}}(u_t, \tau_t, q) \right)
\] 
This final commitment $\pi_{\mathcal{T}}$ represents either the minting or unlocking of tokens within \( \tau_t \). Similar to the source chain process, the token contract on \( \mathcal{T} \) emits a Transfer or Mint event. The Transfer event can be represented as 
\[
\epsilon_{\tau_t, \mathcal{T}} = \emph{Transfer}(\emph{bridge}, u_t, q)
\] 
where tokens are being unlocked (i.e., transferred from the bridge contract). The bridge contract also emits an event accordingly:
\[
\epsilon_{\emph{bridge}, \mathcal{T}} = TokenDeposited(\emph{deposit\_id}, \pi_{\mathcal{T}}(u_t, \tau_t, q)).
\]
It is crucial that this commitment in \( \mathcal{T} \), and subsequent emission of events in both contracts, is only created if \textbf{(1)} \( \phi(\pi_S(u_s, \tau_s, q)) \) holds true, where \( \phi \) is a commitment verification function on \( \mathcal{T} \) that verifies the commitment originating from \( S \); and \textbf{(2)} \( \pi_S \) was created in a transaction, in a block that is \emph{k} blocks deep into blockchain $\mathcal{S}$ and the probability of being reverted is negligible. In this paper, we abstract away the specific implementation details of commitments (e.g., zero-knowledge or Merkle proofs) and focus on the observable state changes. By analyzing state transitions, we can detect anomalies in cross-chain bridges independently of their internal logic, enabling a middleware-level system like \toolName~to reason about cross-chain activity in a protocol-agnostic way.

The withdrawal flow ($\mathcal{T} \rightarrow \mathcal{S}$) is the inverse but very similar, thus not represented.
The key difference is that, usually, the user triggers the final transaction on $\mathcal{S}$, instead of being the validators managing the process (e.g.,~\cite{gnosis_bridge}). This allows the operator to minimize operational costs because validators are not required to issue Ethereum transactions for every withdrawal request.

\subsection{Attacks in Cross-Chain Bridges}


Since June 2021, attackers have stolen more than 3.2 billion USD from cross-chain bridges~\cite{augusto_sok_2024, 10174993}. Hackers target smart contracts that have permission to lock, unlock, mint, or burn tokens. If an attacker gains control of a critical contract -- through a bug or a compromised private key~\cite{augusto_sok_2024} -- they can execute unauthorized token operations. In a lock-unlock model, attackers exploit bridges in two main ways:
\begin{enumerate}
    \item Steal funds held by the bridge contract on \(\mathcal{S}\). Those funds represent the current total value locked by users.
    \item Steal existing funds (liquidity) on \(\mathcal{T}\) that support the unlocking process.
\end{enumerate}
In the burn-mint or lock-mint models, instead of stealing existing funds, attackers mint (create) tokens out of thin air and transfer them to their addresses. These attacks are classified in the literature into two categories based on the direction of the invalid state changes:
\begin{enumerate}
    \item \textbf{Forged Deposit Attack:} Attackers claim funds -- either unlocking existing tokens or minting new ones -- on \(\mathcal{T}\) without locking or burning tokens on \(\mathcal{S}\).
    \item \textbf{Forged Withdrawal Attack:} Attackers withdraw funds on \(\mathcal{S}\) -- similarly, unlocking existing tokens or minting new ones -- without previous burn or lock operations on \(\mathcal{T}\).
\end{enumerate}

%% file: sections/solution.tex
\section{\toolName}
\label{sec: xchainwatcher}

\toolName\footnote{\url{https://github.com/AndreAugusto11/XChainWatcher}}~ is a logic-based \textit{monitoring system} for cross-chain bridges, built as an open-source framework using Souffle~\cite{jordan2016souffle} -- a high-performance Datalog-inspired engine.

\input{listings/Datalog_predicates}

\begin{figure}[ht]
    \centering
    \includegraphics[width=0.49\textwidth]{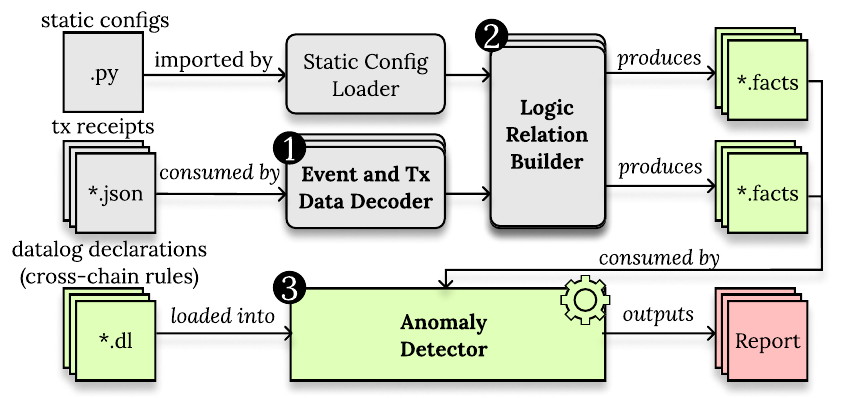}
    \caption{\toolName~relies on event and transaction data decoders, logic relation builders, and a Datalog engine to evaluate relations. The \textit{Decoder} and \textit{Logic Relation Builders} are designed to be pluggable and extensible -- i.e., anyone can extend \toolName~to support other protocols.}
    \label{fig: cross-chain-rules-validator-pipeline}
    \Description[<short description>]{<long description>}
\end{figure}

The workflow of \toolName~is presented in Figure~\ref{fig: cross-chain-rules-validator-pipeline}. There are three phases: 1) decoding event and transaction data from blockchains, 2) building a set of logic relations based on the data extracted, and 3) evaluating relations using a set of detection rules. We design \toolName~to be generic and extensible so that anyone can integrate support for any bridge. Additionally, the logical rules can be fine-tuned for each supported bridge.

\subsection{Logical Relations}\label{subsec: logical_relations}

We model cross-chain operations by defining a comprehensive set of logical relations (i.e., the cross-chain model) that capture events emitted by smart contracts and static configurations common to bridge protocols. These logical relations form the basis for our analysis. We derived them by thoroughly reviewing the open-source code of cross-chain bridge protocols that connect Ethereum to sidechains, and their documentation. We also directly interacted with some protocols and observed the different state changes that occurred -- including Polygon~\cite{polygon_portal}, Ronin~\cite{ronin_bridge}, Omnibridge~\cite{gnosis_bridge}, xDAI Bridge~\cite{xdai_bridge}, and the Nomad Bridge~\cite{nomad_bridge}. These bridges connect Ethereum to multiple sidechains, such as Ronin, Gnosis, Polygon, and Moonbeam. The list of relations (Datalog facts) is in Listing~\ref{listing: datalog_facts}.

\textbf{Contract Events.} The \RuleName{native\_deposit} relation records deposit events of native currency on \(\mathcal{S}\) through the wrapped version of the native currency (e.g., Wrapped Ether contract on Ethereum). The \RuleName{native\_withdrawal} relation logs native token transfers on the target chain when initiating withdrawal of funds (\(\mathcal{T} \rightarrow \mathcal{S}\)), also using the contract representing the wrapped version of the native currency. For bridge-specific events, we use \RuleName{sc\_token\_deposited} and \RuleName{tc\_token\_deposited} to capture token deposits in the bridge contract on the source and target chains, respectively. In parallel, the \RuleName{tc\_token\_withdrew} and \RuleName{sc\_token\_withdrew} relations track token withdrawal events emitted by the bridge contract from the target and source chains. Finally, the \RuleName{erc20\_transfer} relation logs ERC20 token transfers. We also capture mined blockchain transactions through the \RuleName{transaction} relation.

\textbf{Static Configurations.} The \RuleName{bridge\_controlled\_address} relation lists all addresses controlled by the bridge. The \RuleName{token\_mapping} relation links equivalent tokens across chains -- a standard practice in the field~\cite{nomad_docs_token_mapping, polygon_mapped_tokens}. We capture each chain's finality time in the \RuleName{cctx\_finality} relation, modeling the necessary confirmation duration in seconds. Finally, the \RuleName{wrapped\_native\_token} relation identifies wrapped versions of native currencies on each chain -- i.e., the wrapped version of Ether, the native currency of the Ethereum blockchain, is Wrapped Ether (WETH).

\subsection{Decoders and Logic Relation Builders}\label{subsec: anomalies_in_bridges}
The \emph{Static Configuration Loader} imports static facts from the bridge configuration file\footnote{An example of configuration file is at \url{https://github.com/AndreAugusto11/XChainWatcher/blob/main/cross-chain-rules-validator/utils/ronin_env.py}} -- this is information that does not depend on event or transaction data: \RuleName{bridge\_controlled\_address}, \RuleName{token\_mapping}, \RuleName{wrapped\_native\_token}, and \RuleName{cctx\_finality}. On the other hand, the \emph{Event and Transaction Data Decoder} is designed to be bridge-specific, where the remaining relations are extracted from the data decoded from bridge events. This component can be fine-tuned for each bridge allowing the extension of \toolName~to support any protocol.

The input for the latter component is a set of transaction receipts. Each receipt contains the events emitted by all contracts with which the transaction interacted. In many cases, the transaction receipt is sufficient to extract all the facts. In other cases, however, it is not enough, namely, when dealing with native token transfers or when the user uses intermediary protocols to interact with a bridge. In the first case, the \emph{sender} transfers funds natively in a transaction (in the \emph{tx.value} field). In the latter, funds are transferred in internal transactions~\cite{internal_txs}. In both cases, the transferred value is not accessible by the transaction receipt. In this case, we obtain the transaction data by making a request to an RPC node using the RPC methods \emph{eth\_getTransaction} or the \emph{debug\_traceTransaction} with the parameter \emph{\{``tracer'':``callTracer''\}}~\cite{callTracer} that outputs the execution traces and transferred values.

When decoding data and building the logical relations, each transaction is assumed to potentially emit an unlimited number of events, e.g., when batching operations are involved. The extraction of data from relevant events involves extracting the first element in the \textit{topics} list of the transaction receipt, which is equal to the hash of the event signature. For instance, \Keyword{topic[0]} for any event with signature \Keyword{Deposit(address,address,uint256)} is calculated with a hashing function \Keyword{keccak256("Deposit(address,address,uint256)")}.

\subsection{Cross-Chain Rules}\label{subsec: cross_chain_rules}

\textbf{Overview.} We model the expected behavior of bridges (anomaly-based intrusion detection) instead of modeling specific attacks (signature-based intrusion detection) using cross-chain rules~\cite{bace2001intrusion}. This approach allows us to identify anomalies that have not yet been discovered and that are under the hood of the complexity of analyzing cross-chain data. Each rule enforces a set of validations to determine the validity of events within one or more blockchains. Rules are classified as \textit{isolated} (I) or \textit{dependent} (D). An isolated rule concerns only one blockchain, such as the deposit of tokens in $\mathcal{S}$. In contrast, a dependent rule relies on prior state changes on another blockchain, such as the deposit of tokens in $\mathcal{T}$, which depends on tokens being deposited in $\mathcal{S}$. Each rule is also prefixed with SC, TC, or CCTX to indicate whether it is a check on $\mathcal{S}$, $\mathcal{T}$, or both chains, respectively\footnote{The complete definition of all rules in the form of Horn Clauses is in~\url{https://github.com/AndreAugusto11/XChainWatcher/blob/main/cross-chain-rules-validator/datalog/acceptance-rules.dl}}. 

\textit{Rule 1 (I).} \RuleName{SC\_ValidNativeTokenDeposit} ensures a valid deposit of native tokens by the user in $\mathcal{S}$. This rule specifies a relationship between the transaction issued by the user, the event emitted by the bridge contract, and the event emitted by the contract representing the wrapped version of the native currency. In more detail, the checks are: \textbf{(1)} a bridge contract must emit a \emph{Deposit} event; \textbf{(2)} there is a non-reverting transaction that transfers the same amount of tokens natively in \RuleName{tx.value}; \textbf{(3)} there is an event emitted by the token contract asserting the creation of a wrapped version of the native currency through a deposit event \textbf{(4)} the token contract provided is indeed a version of the native currency of $\mathcal{S}$; \textbf{(5)} the validity of the token mappings (i.e., if users are trying to deposit tokens into $\mathcal{T}$ using a different token than what they are using in $\mathcal{S}$); and finally \textbf{(6)} the order of the events emitted by each contract (events emitted by token contracts precede events emitted by bridge contracts -- cf.~Figure~\ref{fig: cross_chain_bridge_model}). In check \textbf{(2)} we do not check whether the transaction targets a bridge contract, as it may target an intermediary protocol contract (e.g., a bridge aggregator~\cite{Subramanian_2024}), which in turn issues an internal transaction to the bridge. We only verify that the deposit event from the token contract must escrow tokens to a valid bridge contract, asserted using \RuleName{bridge\_controlled\_address}. This rule ensures that bridge contracts do not emit events asserting the deposit of tokens if the corresponding value was not effectively sent to the bridge -- and the other way around. An attack that would have been identified using this rule is~\cite{meter_network_hack}. 

\begin{footnotesize}
\begin{Verbatim}[commandchars=\\\{\}]
\textcolor{blue}{SC_ValidNativeTokenDeposit}(...args...) :-
  \textcolor{blue}{sc_token_deposited}(tx_hash, bridge_evt_idx, _, _, dst_token,
                    src_token, dst_chain_id, amount), \textcolor{ForestGreen}{(1)}
  \textcolor{blue}{sc_deposit}(tx_hash, token_evt_idx, sender, bridge_address, amount), \textcolor{ForestGreen}{(3)}
  \textcolor{blue}{transaction}(_, src_chain_id, tx_hash, _, sender, _, amount, 1, _), \textcolor{ForestGreen}{(2)}
  \textcolor{blue}{token_mapping}(src_chain_id, dst_chain_id, src_token, dst_token), \textcolor{ForestGreen}{(5)}
  \textcolor{blue}{wrapped_native_token}(src_chain_id, src_token), \textcolor{ForestGreen}{(4)}
  \textcolor{blue}{bridge_controlled_address}(src_chain_id, bridge_address),
  bridge_evt_idx > token_evt_idx. \textcolor{ForestGreen}{(6)}

\end{Verbatim}
\end{footnotesize}

\textit{Rule 2 (I).} \RuleName{SC\_ValidERC20TokenDeposit} ensures that a valid deposit of ERC20 tokens on the bridge is subject to a series of checks. Specifically, this rule defines a bidirectional relationship \( \epsilon_{\tau_s, S} \Longleftrightarrow \epsilon_{\emph{bridge}, S} \) for ERC20 tokens. This means that whenever a state change involves the transfer of ERC20 tokens, the bridge contract must emit an event corresponding to the commitment described by the initial event, and vice versa. The remaining checks presented in \textit{Rule 1} are also enforced.

Failure to comply with Rule 1 or 2 suggests that a user has deposited tokens in the bridge without the bridge recognizing the deposit. Conversely, if a token transfer event occurs, but no corresponding event is emitted by the bridge contract (or value transferred in the transaction), it could signal an attack, where an attacker bypasses the cross-chain logic and steals funds. Examples of attacks that would have been identified using this rule are \cite{qubit_hack_rekt, multichain_hack_2022_2, THORChain_hack_2021}. Rules 1 and 2 guarantee that the flow 1 -- 4 (in blue) in Figure~\ref{fig: cross_chain_bridge_model} is valid for native and ERC20 tokens, respectively.

\begin{footnotesize}
\begin{Verbatim}[commandchars=\\\{\}]
\textcolor{blue}{SC_ValidERC20TokenDeposit}(...args...) :-
  \textcolor{blue}{sc_token_deposited}(tx_hash, bridge_event_index, _, _, dst_token,
                    src_token, dst_chain_id, amount),
  \textcolor{blue}{erc20_transfer}(tx_hash, src_chain_id, token_event_index, src_token,
                    _, bridge_addr, amount),
  \textcolor{blue}{transaction}(timestamp, src_chain_id, tx_hash, _, from, _, "0", 1, _),
  \textcolor{blue}{token_mapping}(src_chain_id, dst_chain_id, src_token, dst_token),
  \textcolor{blue}{bridge_controlled_addr}(src_chain_id, bridge_addr),
  bridge_event_index > token_event_index.

\end{Verbatim}
\end{footnotesize}

\textit{Rule 3 (I).} \RuleName{TC\_ValidERC20TokenDeposit} outputs valid token deposits in $\mathcal{T}$. It captures the valid relation between the event emitted by the bridge contract and the respective token contract in which tokens are being unlocked/minted. Similarly to \textit{Rules 1 and 2}, there is a bidirectional relationship \( \epsilon_{\tau_t, T} \Longleftrightarrow \epsilon_{\emph{bridge}, T} \). In this instance, tokens are always transferred in the context of a token contract and never natively, thus, we do not need a rule for native token transfers. These events must match variables such as the sender, beneficiary, token, amount being transferred, and order of events. This rule guarantees that flow 5 -- 9 (in green) in Figure~\ref{fig: cross_chain_bridge_model} is valid for any token that is deposited.

\begin{footnotesize}
\begin{Verbatim}[commandchars=\\\{\}]
\textcolor{blue}{TC_ValidERC20TokenDeposit}(...args...) :-
  \textcolor{blue}{tc_token_deposited}(tx_hash, bridge_event_index, deposit_id,
                    beneficiary, dst_token, amount),
  \textcolor{blue}{erc20_transfer}(tx_hash, chain_id, token_event_index, dst_token,
                    bridge_addr_2, beneficiary, amount),
  \textcolor{blue}{transaction}(_, chain_id, tx_hash, _, _, bridge_addr_1, "0", 1, _),
  \textcolor{blue}{bridge_controlled_addr}(chain_id, bridge_addr_1),
  \textcolor{blue}{bridge_controlled_addr}(chain_id, bridge_addr_2),
  bridge_event_index > token_event_index.

\end{Verbatim}
\end{footnotesize}

\textit{Rule 4 (D).} \RuleName{CCTX\_ValidDeposit} correlates events from both \( \mathcal{S} \) and \( \mathcal{T} \), cross-referencing token deposit events across these chains to generate a list of valid \CCTXS. A valid cross-chain transaction for a deposit requires that all parameters from events on both chains be consistent (e.g., token amounts, sender, beneficiary). Furthermore, the causality between these events must be preserved (e.g., the transaction on \( \mathcal{T} \) occurs after the transaction on \( \mathcal{S} \)). Formally, there is a dependency between the commitment on \( \mathcal{T} \) and the commitment on \( \mathcal{S} \), as well as the corresponding events: \( \epsilon_{\emph{bridge}, T} \Longleftarrow \epsilon_{\emph{bridge}, S} \). Since this rule spans multiple blockchains, we must consider their finality times, which we enforce through the \RuleName{cctx\_finality} fact. Failure to comply with this rule indicates, for example, that tokens were moved on only one side of the bridge, such as in the \textbf{Forged Deposit Attack}. This rule would have identified cross-chain hacks such as~\cite{bnb_network_hack, wormhole_hack_rekt, anyswap_hack_2021, poly_network_hack_1, ChainSwap_hack_2021}. This rule guarantees that the entire flow of Figure~\ref{fig: cross_chain_bridge_model} is valid.

\begin{footnotesize}
\begin{Verbatim}[commandchars=\\\{\}]
\textcolor{blue}{CCTX_Deposit}(...args...) :-
  \textcolor{blue}{TC_ValidERC20TokenDeposit}(...args...),
  (
      \textcolor{blue}{SC_ValidERC20TokenDeposit}(...args...) ;
      \textcolor{blue}{SC_ValidNativeTokenDeposit}(...args...)
  ),
  \textcolor{blue}{cctx_finality}(src_chain_id, src_chain_finality),
  \textcolor{blue}{token_mapping}(src_chain_id, dst_chain_id, src_token, dst_token),
  src_chain_ts + src_chain_finality <= dst_chain_ts.

\end{Verbatim}
\end{footnotesize}

We also model the token withdrawal process (\( \mathcal{T} \rightarrow \mathcal{S} \)). Given its similarity to the deposit of tokens, we do not provide a detailed explanation of the related rules. Instead, we briefly overview their goal and definitions. \textit{Rule 5 (I).} \RuleName{TC\_ValidNativeTokenWithdrawal} ensures that native token withdrawals on the target chain \(\mathcal{T}\) are valid. Specifically, a withdrawal must correspond to a \emph{Withdraw} event emitted by the bridge contract and a non-reverting transaction locking or burning funds. This rule is essentially the inverse of Rule 1, applying similar checks but in the withdrawal context. Rule 5 would have identified one attack~\cite{thorchain_hack_2021_3}. \textit{Rule 6 (I).} \RuleName{TC\_ValidERC20TokenWithdrawal} applies to ERC20 token withdrawals on \(\mathcal{T}\), ensuring that any withdrawal event emitted by the bridge contract matches a corresponding \emph{Transfer} event for the ERC20 tokens being withdrawn. This rule is analogous to Rule 2, and would have identified one attack~\cite{thorchain_rekt_2021_2}. \textit{Rule 7 (I).} \RuleName{SC\_ValidERC20TokenWithdrawal} extends these checks to ERC20 withdrawals on the source chain \(\mathcal{S}\), mirroring Rule 3’s checks in the reverse direction. Finally, \textit{Rule 8 (D).} \RuleName{CCTX\_ValidWithdrawal} links withdrawal events on \(\mathcal{T}\) and \(\mathcal{S}\), verifying that all parameters across the chains match and enforcing the correct causal relationship between events, similar to Rule 4 for deposits but in reverse. Rule 8 would have identified multiple attacks such as the \textbf{Forged Withdrawal Attack}~\cite{ronin_hack_rekt, poly_network_hack_2, bxh_hack_2021, multichain_hack_2023, pnetwork_hack}.

While it is impossible to design generic rules that allow for every existing bridge, we highlight that these rules can be easily extended/fine-tuned to find anomalies in other bridges.

%% file: listings/datalog_predicates.tex
\lstset{style=datalog}

\begin{lstlisting}[language=Datalog, caption={Definition of the logical relations built by \toolName.}, label={listing: datalog_facts}, float=*]
.decl native_deposit(tx_hash, chain_id, event_index, from, to, amount).
.decl native_withdrawal(tx_hash, chain_id, event_index, from, to, amount).
.decl sc_token_deposited(tx_hash, event_index, deposit_id, beneficiary, dst_token, orig_token, dst_chain_id, amount).
.decl tc_token_deposited(tx_hash, event_index, deposit_id, beneficiary, dst_token, amount).
.decl tc_token_withdrew(tx_hash, event_index, withdrawal_id, beneficiary, orig_token, dst_token, dst_chain_id, amount).
.decl sc_token_withdrew(tx_hash, event_index, withdrawal_id, beneficiary, dst_token, amount).
.decl erc20_transfer(tx_hash, chain_id, event_index, contract, from, to, amount).
.decl transaction(timestamp, chain_id, tx_hash, from, to, value, status, fee).
.decl bridge_controlled_address(chain_id, bridge_address).
.decl token_mapping(source_chain_id, target_chain_id, source_chain_token, target_chain_token).
.decl cctx_finality(chain_id, finality_seconds)
.decl wrapped_native_token(chain_id, token).
\end{lstlisting}

%% file: sections/evaluation.tex
\section{Evaluation Methodology}\label{sec: evaluation}

We evaluate \toolName~using the cross-chain rules presented in the last Section and detail the anomaly detection analysis in the Ronin and Nomad bridges.

\subsection{Data Sources}

We selected two previously exploited bridges to analyze the capabilities of \toolName~and the cross-chain rules: the \textbf{Nomad bridge} and the \textbf{Ronin bridge}. This selection allows us to test \toolName~against bridges that have suffered attacks and whose architecture and security assumptions differ (\S\ref{subsubsec: nomad} and \S\ref{subsubsec: ronin}). We used Blockdaemon's Universal API~\cite{blockdaemon_universal_api} to retrieve blockchain data from the Ethereum mainnet. We implemented a fallback to native RPC methods when the API could not provide the necessary data (namely \textit{eth\_getLogs} and \textit{eth\_getTransactionReceipt}). Additionally, we used these methods to extract data from Moonbeam and Ronin blockchains that are not supported by the API. We gathered addresses of interest, including various versions of deployed contracts through documentation and analysis of the source code of each bridge\footnote{an example for the Nomad bridge is in \url{https://anonymous.4open.science/r/XChainWatcher-B5F1/cross-chain-rules-validator/utils/nomad_env.py}}.

\subsubsection{Time Frames}

Since we adopt an anomaly-based intrusion detection approach (instead of signature-based), which tends to have a high false positive rate~\cite{bace2001intrusion}, we choose to evaluate protocols over smaller time frames. This approach enables us to analyze each flagged anomaly individually, determining whether it results from a modeling error or represents a previously unidentified anomaly in cross-chain bridges. Additionally, we want to study particular attacks and their consequences -- involving bigger timeframes would involve significantly more data, without necessarily providing additional relevant insights.
Table~\ref{table: timestamps} lists the timestamps used for data extraction. We select an interval of interest for both bridges that includes the attack dates, denoted $[t_1;t_2]$. To avoid missing cross-chain transactions occurring near the start and end of the interval of interest, we incorporate additional intervals before and after that interval ($[t_0,t_1[$ and $]t_2,t_3]$). This is relevant, for example, when there is a deposit of tokens in $\mathcal{S}$ near $t_2$ and the corresponding transaction in $\mathcal{T}$ falls outside $[t_1;t_2]$ (within $]t_2;t_3]$).

\input{tables/time-intervals}

To analyze the Nomad bridge, we extracted 20,551 transactions from Ethereum, 16,737 transactions from Moonbeam, and 20,308 transactions from/to other blockchains, which were only used for data analysis. In the additional period, we collected additional 1,774 transactions on Ethereum, from the latest versions of the deployed bridge contracts. 
On the Ronin bridge, we extracted 72,820 transactions from Ethereum and 75,102 from Ronin. In the additional period, we collected additional 516,657 and 151,325 transactions on Ethereum and Ronin, respectively. The data collection totaled 875,274 transactions across the studied bridges and blockchains. 

\subsubsection{Nomad Bridge}\label{subsubsec: nomad}

The Nomad bridge supports six blockchains. We select the most active blockchains in terms of bridge usage: Ethereum ($\mathcal{S}$) and Moonbeam ($\mathcal{T}$). The bridge operates based on fraud proofs \cite{nomad_bridge} -- i.e., a set of relayers transfers state proofs between blockchains, and the watchers (which are off-chain parties) have a predefined time window to challenge the relayed data. The data is optimistically accepted if no challenge is received within this window. According to the project documentation, this time window was set to $30$ minutes~\cite{nomad_optimistic_period} during the selected time frame. The main bridge contract on Moonbeam was deployed on January 11, 2022 ($t_1$). Since we start our analysis on this date, there is no $t_0$ ($t_0 = t_1$). The Nomad bridge was exploited on August 2, 2022, causing the bridge contracts to be paused until December 15, 2022. After this date, new transactions depositing tokens into the bridge on Ethereum started being reissued. 
In $]t_2;t_3]$ we only collect withdrawals in Ethereum to match all the withdrawal requests performed on Moonbeam in $[t_1;t_2]$ that did not complete.

\subsubsection{Ronin Bridge}\label{subsubsec: ronin}

The Ronin bridge connects Ethereum ($\mathcal{S}$) and the Ronin blockchain ($\mathcal{T}$) and operates based on a multi-signature of trusted validators~\cite{ronin_bridge} -- i.e., deposits and withdrawals are executed when a threshold of validators attests the validity of the action on the origin blockchain (be it a lock or burn of tokens). The Ronin bridge was deployed in early 2021, and the attack occurred on March 22, 2022. The interval of interest spans approximately four months, from the start of 2022 to April 28, 2022, when the main bridge contract on Ronin was paused (\AddrHref{https://app.roninchain.com/tx/0xe806b36b9f337e8512dd806a5845451232a0da52c66f2921c4f7e222bd5e19fd}{0xe806b36b9f337e8512dd806a5845451232a0da52c66f2921c4f7e222bd5e19fd}). To capture incomplete withdrawals on $\mathcal{T}$ before the attack, we analyze additional data on Ethereum between $]t_2;t_3]$. This required scanning for events in the newer version of the main bridge contract (\AddrHref{https://etherscan.io/address/0x64192819ac13ef72bf6b5ae239ac672b43a9af08}{0x64192819ac13ef72bf6b5ae239ac672b43a9af08}), which was deployed on Ethereum after the attack on June 22, 2022. Finally, based on the same logic as above, we also captured additional deposits in Ethereum to capture cross-chain transactions initiated in $[t_0;t_1[$, whose deposit in Ronin is at the beginning of $[t_1;t_2]$.

\subsection{Experiment Setup}

 



We present the performance analysis of \toolName, using the rules defined in Section~\ref{subsec: cross_chain_rules}, in finding anomalies in the Nomad and Ronin bridges. We divide the analysis into two main processes: 1) decoding data and building the Datalog facts, and 2) running the cross-chain rules to find anomalies. We computed the results on a MacBook Pro with a 14-core M3 Max processor and 36GB of RAM.

\subsubsection{Decoding and Extracting Data}

\begin{figure}[ht]
    \centering
    \includegraphics[width=0.5\textwidth]{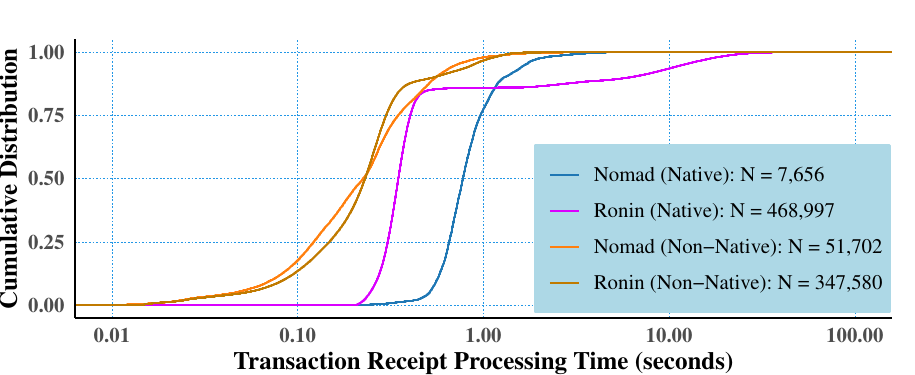}
    \caption{Cumulative distribution of transaction receipt processing time, reflecting the latency of extracting all facts from a transaction receipt for native and non-native token transfers.}
    \label{figure: ronin-evaluation}
    \Description[<short description>]{<long description>}
\end{figure}

Figure~\ref{figure: ronin-evaluation} illustrates the cumulative distribution of transaction receipts processing time, differentiating between transfers of native and non-native funds in each bridge. Additional metrics are provided in Table~\ref{table: evaluation}. Transactions transferring native tokens take longer because the transaction receipt is not enough to get \Keyword{tx.value}, thus requiring at least one extra time-consuming RPC call. Furthermore, for native value transfers, some transactions exhibited unusually high latencies (e.g., $6.5\%$ exceeded $10$ seconds, with one instance reaching $138.15$ seconds). This delay mainly results from the high latency of the \emph{debug\_traceTransaction} method when making RPC requests to an Ethereum node~\cite{rpc_json_ethereum}. Not only is this a resource-intensive method, but multiple timeouts caused various retries to retrieve the data. A more stable RPC node connection -- ideally hosting one alongside \toolName~-- and extending the timeout period for these resource-intensive methods would significantly reduce the latency, dropping towards the median ($0.35$ and $0.78$ seconds, for Ronin and Nomad, respectively).

\input{tables/evaluation}

\subsubsection{Executing the Cross-Chain Rules}

Based on the data extracted, we run the detection rules to identify anomalies. In addition to the rules presented in Section~\ref{sec: cross-chain-bridge-model}, we implemented additional Datalog rules to compare datasets and perform a more fine-grained analysis -- we created 30 logical rules in total, available in the linked repository. The total time consists of decoding the data and building the logic relations plus the execution of the detection rules. 
For the Ronin bridge, the model processed more than 1,570,000 data tuples, producing results, on average, in $3.58$ seconds, while for the Nomad bridge, it analyzed more than 200,000 data tuples and generated results in $0.51$ seconds\footnote{detailed results can be found in \url{https://anonymous.4open.science/r/XChainWatcher-B5F1/profiler_html/ronin.html}}. 


\subsubsection{Preliminary Findings of Cross-Chain Transactions}\label{subsubsec: cctx_latency}

\begin{figure}[ht]
    \centering
    \includegraphics[width=0.45\textwidth]{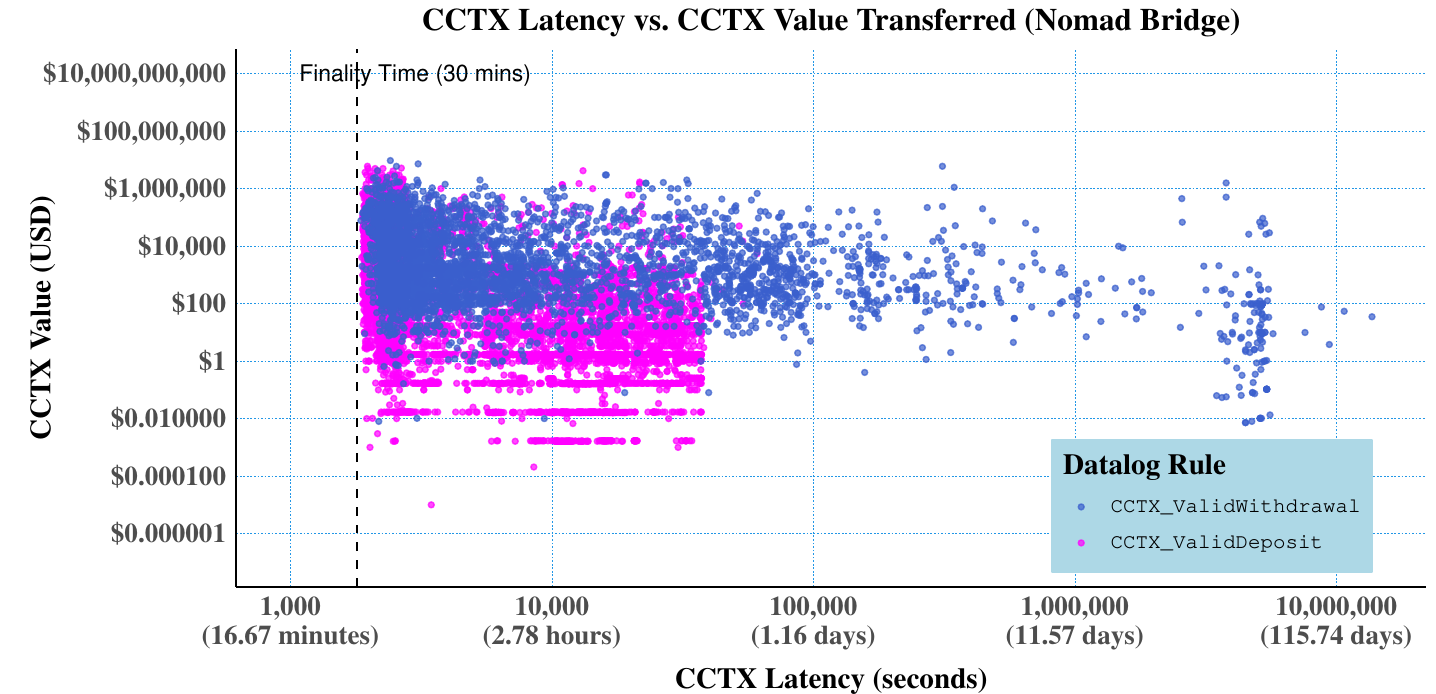}
    \caption{Correlation between the latency and value transferred in each cctx completed before the attack.
    }
    \label{figure: cctx_latency_vs_value_nomad}
    \Description[<short description>]{<long description>}
\end{figure}

A byproduct of our work is a dataset of cross-chain transactions captured by rules 4 and 8 -- i.e., data from two blockchains that are linkable and represent valid cross-chain token transfers. Figure~\ref{figure: cctx_latency_vs_value_nomad} presents the latency associated with each cross-chain transaction identified on the Nomad bridge (the Ronin bridge data was omitted for the sake of space but provides the same insights).
We call out two main insights: 1) the dispersion of the latency of withdrawals is much higher, which is due to the users being the ones responsible for issuing the final transaction on the destination blockchain, contrary to the deposit process (cf. Section~\ref{sec: cross-chain-bridge-model}) -- the slowest \CCTX~took more than 5 months to complete (\AddrHref{https://moonscan.io/tx/0x8afeeea543a4516c279bff2748b3bbede9cc916cc535524d62433368119a85bb}{0x8afeeea543a4516c279bff2748b3bbede9cc916cc535524d62433368119a85bb} in $\mathcal{T}$ and \AddrHref{https://etherscan.io/tx/0xdfaaeecb7f0dda43f02966997039b2b75169a3faa3ae5063d74a348ceb98e3cb}{0xdfaaeecb7f0dda43f02966997039b2b75169a3faa3ae5063d74a348ceb98e3cb} in $\mathcal{S}$); 2) all \emph{cctx}s identified by \toolName~start at the 30-minute mark, which aligns with our expectations, as the Nomad fraud-proof window is set for this period enforced by \RuleName{cctx\_finality};

\input{tables/anomaly-detection-analysis}

\section{Anomaly Detection Results}\label{sec: results}
Hereafter, we present the results of the anomaly detection rules. Table~\ref{table: anomaly_detection_results} shows the number of detected anomalies and the reasons behind each one. Section~\ref{subsubsec: isolated_rules} discusses the anomalies found by \textit{isolated} rules (1-3 and 5-7), and Section~\ref{subsubsec: dependent_rules} presents and discusses the anomalies found by \textit{dependent} rules (4 and 8).


\subsection{Isolated Rules (Rules 1-3 and 5-7)}\label{subsubsec: isolated_rules}
We start by analyzing the anomalies detected by rules 1-3 and 5-7.

\subsubsection{Depositing on $\mathcal{S}$}
On the Nomad bridge, we detected \SCDepositFactsNomad native value transfers (\RuleName{sc\_deposit}), \ERCTransferSDepositsFactsNomad token deposits (\RuleName{erc20\_transfer}), and \SCTokenDepositedFactsNomad TokenDeposited events emitted by the bridge contract (\RuleName{sc\_token\_deposited}), which reveals that $39$ value transfers did not have a corresponding bridge event emitted. Further analysis showed that $14$ of these transactions are phishing attempts, characterized by numerous events emitted by tokens marked on block explorers as having a bad reputation (e.g., \AddrHref{https://etherscan.io/tx/0x88fc3c5e05aae4d898fc92eb93c64ee71dcbbb2a4e5e3715f994adcbce72864a}{0x88fc3c5e05aae4d898fc92eb93c64ee71dcbbb2a4e5e3715f994adcbce72864a}). The remaining 25 transactions were single-event transactions that called the \textit{Transfer} function of multiple reputable ERC20 tokens, with a total of approximately \textbf{\$93.86K} sent to the bridge without triggering a cross-chain transfer (e.g., \AddrHref{https://etherscan.io/tx/0x7e4e62f98d4c3194e5b3fbef79cf5fda3330287d489dffd6252634f3f6208d88}{0x7e4e62f98d4c3194e5b3fbef79cf5fda3330287d489dffd6252634f3f6208d88}). On the Ronin bridge, we identified $83$ unmatched value transfers, in which $3$ were related to phishing attempts and 80 were also random transfers of value to the bridge contract, which accounts for \textbf{\$113.00K} (e.g., \AddrHref{https://etherscan.io/tx/0xe898f40fa2fe5c8d89df3a2e4f2496bd11daceedce1523afabfbf144b32d148d}{0xe898f40fa2fe5c8d89df3a2e4f2496bd11daceedce1523afabfbf144b32d148d}).

\begin{anomalybox}
\textbf{Finding 1.} Attackers use low-value tokens, usually with the name of known tokens, to interact with bridge contracts. These practices are considered phishing attacks, in which users can be misled into using fake tokens to increase their trading value.

\textbf{Finding 2.} \textbf{Over \$206K} worth of reputable ERC-20 tokens were sent directly to bridge addresses without using protocol contracts. Despite warnings from DeFi platforms about potentially irreversible losses, this risky behavior appears common.
\end{anomalybox}

\subsubsection{Depositing on $\mathcal{T}$}
We found no anomaly in the process of depositing tokens in $\mathcal{T}$.

\subsubsection{Withdrawals on $\mathcal{T}$}\label{feat: withdrawals_on_t}
In the Nomad bridge, we identified three transactions accepted by the bridge where funds were withdrawn to unintended Ethereum addresses 
due to being wrongly formatted -- they interacted with the Nomad bridge contract using a 32-byte string instead of a 20-byte Ethereum address in the \textit{beneficiary} address field.
Therefore, this leads to unparseable data from our tool (the parser is programmed to parse only valid 20-byte addresses). We further discuss this anomaly in Section~\textit{\ref{subsubsec: invalid_benef_addresses} Invalid Beneficiary Addresses}. Beyond these three anomalies, we discovered seven transactions from a single address attempting to exploit the bridge using different inputs in the ``token'' field. The attacker first attempted to provide the address of a malicious smart contract as a token, probably to gain control over the bridge (\AddrHref{https://moonscan.io/tx/0x56e6c554169c0b6e99d744416c04c11926c3a867ae2ffd3125aa5ba0eaf6afe1}{0x56e6c554169c0b6e99d744416c04c11926c3a867ae2ffd3125aa5ba0eaf6afe1}). In the following 3 transactions, the attacker tried to withdraw funds using a newly created contract that was not mapped to a token in $\mathcal{S}$ (e.g., \AddrHref{https://moonscan.io/tx/0xebd68eaaa20de3066cf3f53c26777c38d62251ca13c9d6e0d3a991e011babfa9}{0xebd68eaaa20de3066cf3f53c26777c38d62251ca13c9d6e0d3a991e011babfa9} with token \AddrHref{https://moonscan.io/address/0x24229bf80425c27DB54fB3E4340251Dd5C16Aefb}{0x24229bf80425c27DB54fB3E4340251Dd5C16Aefb}), in an attempt to have tokens minted on $\mathcal{S}$. Finally, in the latter two, the user attempted to withdraw funds from a (fake) token contract called \textit{Wrapped ETH} (\AddrHref{https://moonscan.io/address/0xcbb4825CF7Cf72a88d1BDdd494c1A251CF21b91F}{0xcbb4825CF7Cf72a88d1BDdd494c1A251CF21b91F}), to unlock real ETH on Ethereum (e.g., \AddrHref{https://moonscan.io/tx/0x7cd7d1a4feceeaa14b6c347488229707fa710daeab2e1e5d707d43a720a703f1}{0x7cd7d1a4feceeaa14b6c347488229707fa710daeab2e1e5d707d43a720a703f1}). Fortunately, these transactions reverted and all attempts failed. On the Ronin bridge, we identified two events emitted by the bridge contract without a match on \RuleName{erc20\_transfer} or \RuleName{sc\_withdrawal}. These were trying to withdraw unmapped tokens from $\mathcal{T}$ to $\mathcal{S}$, and therefore no tokens were moved, even though the bridge emitted a \textit{Withdraw} event.

\begin{anomalybox}
\textbf{Finding 3.} Attackers interact with bridge contracts providing fake tokens with symbols or names equal to reputable tokens, in an attempt to deceive the bridge to unlock real funds on the destination blockchain.
\end{anomalybox}

\subsubsection{Withdrawals on $\mathcal{S}$}
The analysis highlights $3$ events where funds were transferred from a bridge address without emitting corresponding bridge events: $2$ in Nomad and $1$ in the Ronin bridge. These instances were linked to phishing attempts and marked accordingly in block explorers (e.g., \AddrHref{https://etherscan.io/tx/0x358788e319dcd2a0afd03102cc944ffda0bf6a68abbd2d84734affcb07d739ca}{0x358788e319dcd2a0afd03102cc944ffda0bf6a68abbd2d84734affcb07d739ca} and \AddrHref{https://etherscan.io/tx/0x78b643a338afa5bd56aa2eeccd5c2381a5e5a921986a6ccf7b678958b7d62766}{0x78b643a338afa5bd56aa2eeccd5c2381a5e5a921986a6ccf7b678958b7d62766}).


\subsection{Dependent Rules (Rules 4 and 8)}\label{subsubsec: dependent_rules}

\input{tables/deposits_and_withdrawals_anomalies}

Now, we analyze the results of the \textit{Dependent} rules (4 and 8). Recall from Section~\ref{subsec: cross_chain_rules} that Rules 4 and 8 capture linked state changes across blockchains -- i.e., for a record to be accepted by these rules, there must be a set of events on both sides of the bridge that are matched. In addition, \RuleName{cctx\_finality} and \RuleName{token\_mapping} must be guaranteed. For example, there may be a valid deposit of tokens in \(\mathcal{S}\) captured by \RuleName{SC\_ValidERC20ValidDeposit}. However, no correspondence is found on \(\mathcal{T}\), which signals that the protocol is not working as intended (e.g., no availability). In this case, the record of \RuleName{SC\_ValidERC20ValidDeposit} is said to be ``unmatched'', since it did not match any event on the other blockchain that complies with Rule 4 \RuleName{CCTX\_Deposit}. Table~\ref{table: cctx_analysis} dissects the anomalies detected in Table~\ref{table: anomaly_detection_results} for \RuleName{CCTX\_ValidDeposit} and \RuleName{CCTX\_ValidWithdrawal}. As an example, Table~\ref{table: anomaly_detection_results} shows that 19 anomalies have been detected using \RuleName{CCTX\_ValidDeposit}. Table~\ref{table: cctx_analysis} clarifies that 6 of these anomalies are deposits of tokens on \(\mathcal{S}\) that did not have a correspondence on \(\mathcal{T}\), and 13 are the opposite -- deposits of tokens on \(\mathcal{T}\) that did not have any prior correspondence on \(\mathcal{S}\).



\subsubsection{Cross-Chain Finality Violations}\label{subsubsec: cctx_finality}
\begin{figure}
    \centering
    \includegraphics[width=0.48\textwidth]{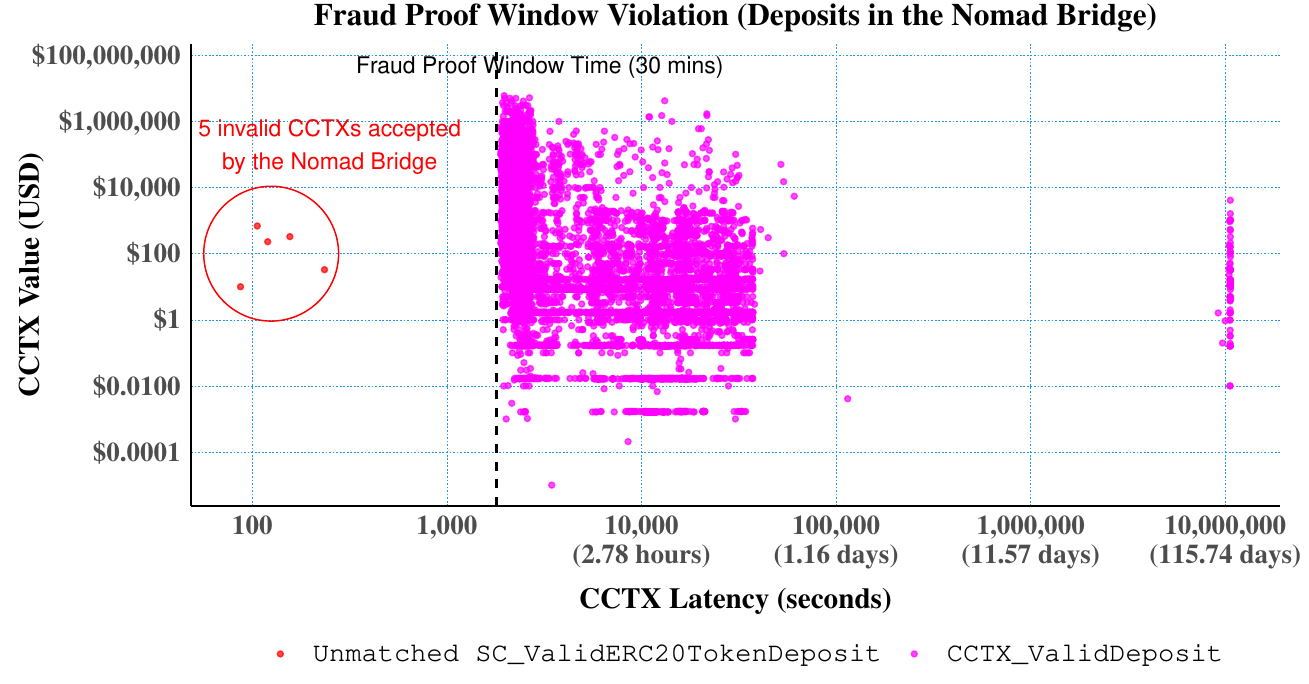}
    \caption{Cross-chain finality violation in the Nomad bridge -- we identified 5 unmatched events on both \FootnotesizeKeyword{SC\_ValidERC20TokenDeposit} and \FootnotesizeKeyword{TC\_ValidERC20TokenDeposit} not captured by \FootnotesizeKeyword{CCTX\_ValidDeposit}, due to non-compliance with the fraud-proof time window.}
    \label{figure: finality_break_nomad}
    \Description[<short description>]{<long description>}
\end{figure}

One of the most intriguing findings in this paper is the identification of $37$ violations of cross-chain rules -- $5$ on the Nomad bridge and $32$ on the Ronin bridge -- which were accepted by both bridges at the time, transferring a total value of \textbf{\$1.3K} and \textbf{\$667K}, respectively.
In the Nomad bridge, $5$ instances from~\RuleName{SC\_ValidERC20TokenDeposit} and $5$ instances from \RuleName{TC\_ValidERC20TokenDeposit} matched each other but were not captured by \RuleName{CCTX\_ValidDeposit} -- i.e., even though there were valid commitments on both sides of the bridge, \toolName~did not consider this a valid deposit. Similarly, on the Ronin bridge, $10$ events were emitted on each side that did not comply with a valid deposit (failed \RuleName{CCTX\_ValidDeposit}), and $22$ events on each side that did not comply with a valid withdrawal (\RuleName{CCTX\_ValidWithdrawal}). Figure~\ref{figure: finality_break_nomad} for the Nomad bridge 
demonstrates why these events were not captured by \RuleName{CCTX\_ValidDeposit} and \RuleName{CCTX\_ValidWithdrawal}.
When depositing tokens using Nomad, in the fastest \CCTX, the time difference between the initial deposit in $\mathcal{S}$ (\AddrHref{https://etherscan.io/tx/0xeb06aa1e251555ac1e4f58b04987d37f87cf407266a5b528f6de235a45590fea}{0xeb06aa1e251555ac1e4f58b04987d37f87cf407266a5b528f6de235a45590fea}) and the corresponding deposit on $\mathcal{T}$ (\AddrHref{https://moonscan.io/tx/0x2cdc80f24ae1c65b88d956c5709514269c76a911002fca7d1efc7cb87e84ef0c}{0x2cdc80f24ae1c65b88d956c5709514269c76a911002fca7d1efc7cb87e84ef0c}) was as short as $87$ seconds, approximately 20 times less than the required fraud-proof window. This finding is particularly concerning because it implies that the security mechanisms of the bridge were bypassed. Not only did it fail to comply with the fraud-proof time window, but it was very close to the finality period of the source chain (Ethereum) at the time of the attack -- before ``The Merge''~\cite{the_merge} was around $78$ seconds. On the Ronin bridge, the fastest deposit took 66 seconds (\AddrHref{https://etherscan.io/tx/0x468868506b014b5729f9926ff8bce17823842747828d9a82b23767a5b408cdf3}{0x468868506b014b5729f9926ff8bce17823842747828d9a82b23767a5b408cdf3} and \AddrHref{https://app.roninchain.com/tx/0xc2997f0a2e14c7db69cafbc6e58839299d8130a4c3310e49dbf5fb62f707279d}{0xc2997f0a2e14c7db69cafbc6e58839299d8130a4c3310e49dbf5fb62f707279d}), which was less than Ethereum's finality period. However, the fastest withdrawal took 11 seconds ($11 < 45$, where $45$ seconds was Ronin's finality period at the time). These practices pose a considerable risk to \CCTX~ validation, creating multiple potential attack vectors, particularly for smaller blockchains or those more susceptible to forks. 

\begin{anomalybox}
\textbf{Finding 4.} We identified 37 instances where the protocol-defined finality was not satisfied. In Nomad, this was due to smart contract enforcement issues of the fraud-proof window; in Ronin, off-chain validators failed to enforce the source chain’s finality period.
\end{anomalybox}

\subsubsection{Invalid Beneficiary Addresses}\label{subsubsec: invalid_benef_addresses}
In Nomad, users must specify a beneficiary address when transferring funds. To accommodate multiple destination blockchains, Nomad uses a 32-byte field for the beneficiary address instead of a 20-byte address, since some blockchains (e.g., Solana) require 32 bytes. When transferring funds to an EVM-based blockchain, users must left-pad the address with zeros, and the bridge contract extracts the last 20 bytes.

We identified an anomaly when a user submitted a transaction (\AddrHref{https://etherscan.io/tx/0x794135750db90cf346b08dc3de668cb19ea69f59bc59e7f158759508ed9a1393}{0x794135750db90cf346b08dc3de668cb19ea69f59bc59e7f158759508ed9a1393}) in $\mathcal{S}$ that deposited 10 DAI into a beneficiary address that was right padded instead of left-padded. 
The contract extracted the last 20 bytes (mainly \textit{0s}) and expected a left-padded address; our tool, which accepted both left and right padding, parsed the address ``correctly'', i.e., without the padding. The user provided an incorrect input. However, we could not determine whether the error resulted from user misuse or a malfunction in the bridge's UI.

We also detected three anomalies when withdrawing funds in $\mathcal{S}$ (e.g., \AddrHref{https://moonscan.io/tx/0xfcc6d0775cb1cbb2dc4654b563a9b9881b5972e1dd213b0e8b7535bd5b8e7c5f}{0xfcc6d0775cb1cbb2dc4654b563a9b9881b5972e1dd213b0e8b7535bd5b8e7c5f}). These involved events that we could not decode earlier (in Section~\ref{feat: withdrawals_on_t}) because the destination Ethereum address is represented as an unpadded 32-byte string, and therefore represents an invalid Ethereum address. Again, the bridge contract simply extracted the last 20 bytes, whereas our tool throws an error. Interestingly, none of the destination addresses extracted by the bridge contract showed any activity after these transactions, which reveals that the addresses computed by the bridge were not the ones intended by the users -- i.e., users mistakenly provided a wrongly formatted address and lost the funds because they do not control these addresses. 
While these 4 cases can be considered \textbf{false positives} from our tool -- i.e., not protocol anomalies -- they still revealed genuine anomalies in user behavior.

\begin{anomalybox}
\textbf{Finding 5.} Protocols do not safeguard users against incorrectly formatted inputs, as bridge contracts are often designed to be blockchain-agnostic and may lack strict input validation.
\end{anomalybox}

\subsubsection{Invalid Token Mappings}\label{subsubsec: invalid_token_mappings}
We identified 9 anomalies in Rules 4 and 8 due to records not complying with the \RuleName{token\_mapping} predicate, 7 when depositing tokens in \(\mathcal{T}\) using the Nomad bridge, and 2 when withdrawing tokens in \(\mathcal{S}\).

According to the Nomad bridge documentation~\cite{nomad_docs_token_mapping}, anyone can deploy a new token on Moonbeam and ask the bridge to link it to the contract that represents the same token on Ethereum. We found 5 transactions that involved the Nomad bridge operator deploying new ERC20 tokens on Moonbeam. 
One of the transactions \AddrHref{https://moonscan.io/tx/0x7fe7e6ea905831d135514fd665d9867349b24134f0dd1217fb7d55a88204bf27}{0x7fe7e6ea905831d135514fd665d9867349b24134f0dd1217fb7d55a88204bf27} deployed a new token contract on Moonbeam mapped to a token contract in $\mathcal{S}$ called WRAPPED GLMR (e.g., \AddrHref{https://etherscan.io/address/0x92C3A05B5CC7613E8A461968AD8616BAE3C47178}{0x92C3A05B5CC7613E8A461968AD8616BAE3C47178}). This token is the native token of the Moonbeam blockchain, which already exists in $\mathcal{S}$ and is already mapped by the bridge (\AddrHref{https://etherscan.io/address/0xba8d75BAcCC4d5c4bD814FDe69267213052EA663}{0xba8d75BAcCC4d5c4bD814FDe69267213052EA663}) -- and therefore, this mapping should not have been validated by the bridge operator. Our hypothesis is that these may be users creating fake tokens with the name of real tokens (e.g., WRAPPED GLMR), in an attempt to later on withdraw real funds on Ethereum. This vulnerability was the cause of an attack on the Thorchain bridge in 2022~\cite{zhang_xscope_2022}, where attackers created a fake contract called \textit{Wrapped Ether} and tricked the bridge contract into accepting the withdrawal of real Ether. The 4 subsequent transactions tried depositing different amounts of different tokens to multiple addresses in $\mathcal{T}$. Strangely, no activity was found on $\mathcal{S}$ in the mapped contracts -- i.e., the tokens were never used by anyone previously (e.g., \AddrHref{https://moonscan.io/tx/0xda3f048e50e8e4df1d5726fb3ea6839e95ed15e49b4d7daf3c91a5b44b3f5c72}{0xda3f048e50e8e4df1d5726fb3ea6839e95ed15e49b4d7daf3c91a5b44b3f5c72}).
This activity is very unusual, especially since the transactions mapping tokens across blockchains were issued by the Nomad bridge operator, which suggests a lack of contract verification between blockchains by the operator.

\begin{anomalybox}
\textbf{Finding 6.} The Nomad bridge operator linked fake or duplicate tokens between Moonbeam and Ethereum, including an already existing mapping for WRAPPED GLMR. This highlights a lack of rigorous token contract verification, leaving the protocol vulnerable to spoofing attacks.
\end{anomalybox}


\subsubsection{Withdrawals in \texorpdfstring{$\mathcal{T}$}{T} with no Correspondence in \texorpdfstring{$\mathcal{S}$}{S}}
\label{subsec: withdrawals_in_t}

\begin{figure}
    \centering
    \includegraphics[width=0.5\textwidth]{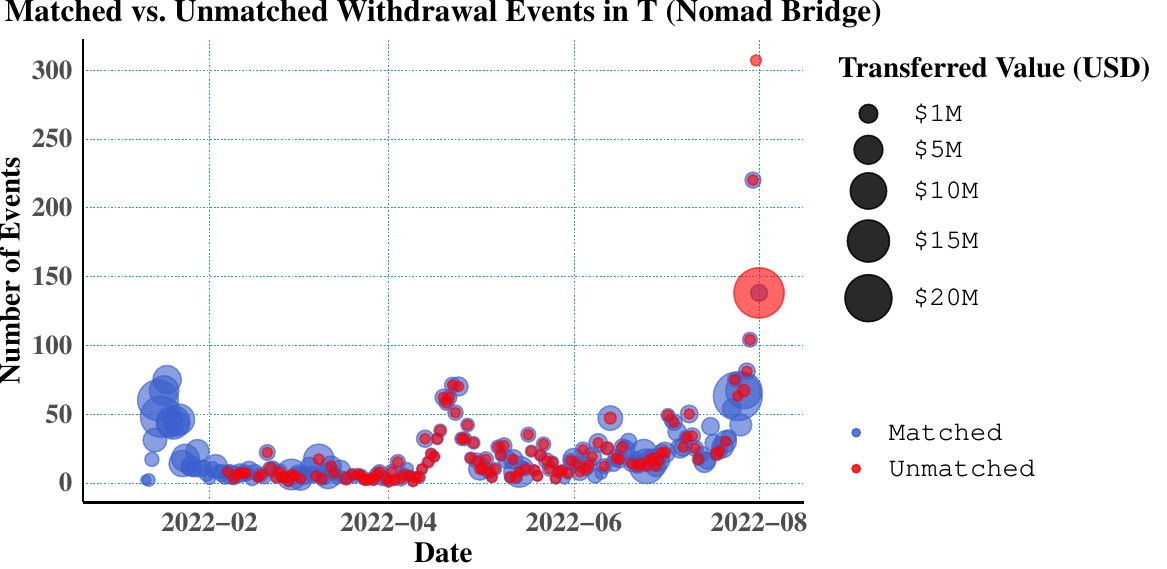}
    \caption{Withdrawal events emitted on $\mathcal{T}$ matched (N = 4,482) or unmatched (N = 828) with another event on $\mathcal{S}$ (through \FootnotesizeKeyword{CCTX\_ValidWithdrawal}) in the Nomad bridge. 
    }
    \label{figure: matched_vs_unmatched_withdrawals_on_tc_nomad}
    \Description[<short description>]{<long description>}
\end{figure}

We found 729 (= 238 + 491) withdrawals on $\mathcal{T}$, in which no corresponding transaction was found in $\mathcal{S}$. A first hypothesis to explain the high number of anomalies is whether these values are a consequence of the attack, i.e., multiple users tried (unsuccessfully) to withdraw funds as the bridge was paused. To test this hypothesis, Figure~\ref{figure: matched_vs_unmatched_withdrawals_on_tc_nomad} shows the comparison between the matched and unmatched withdrawal events emitted on $\mathcal{T}$ on the Nomad bridge (the results for the Ronin bridge are similar but not shown for the sake of space). 
As expected, close to when the hack happened in August 2022, there were many unmatched withdrawal events emitted in $\mathcal{T}$ -- there were $313$ events trying to withdraw $\$24.7$M worth of tokens in the 24 hours prior to the attack. In Ronin, we identified $468$ events withdrawing $\$24.3$M in the same period. Not surprisingly, in the event of an attack, it is difficult to withdraw tokens, due to the bridge being paused after the attack. However, it is also noticeable that throughout the entire period in which the bridge was functioning, there were always multiple low-value unmatched events, following the same trend as the matched ones. These are funds escrowed in $\mathcal{T}$ in which the corresponding tokens were never unlocked on $\mathcal{S}$ within $[t_1;t_2]$ -- and the bridge still holds the escrowed assets in $\mathcal{T}$.

\input{tables/destination-addresses-metrics}

\begin{figure}[ht]
    \centering
    \includegraphics[width=0.5\textwidth]{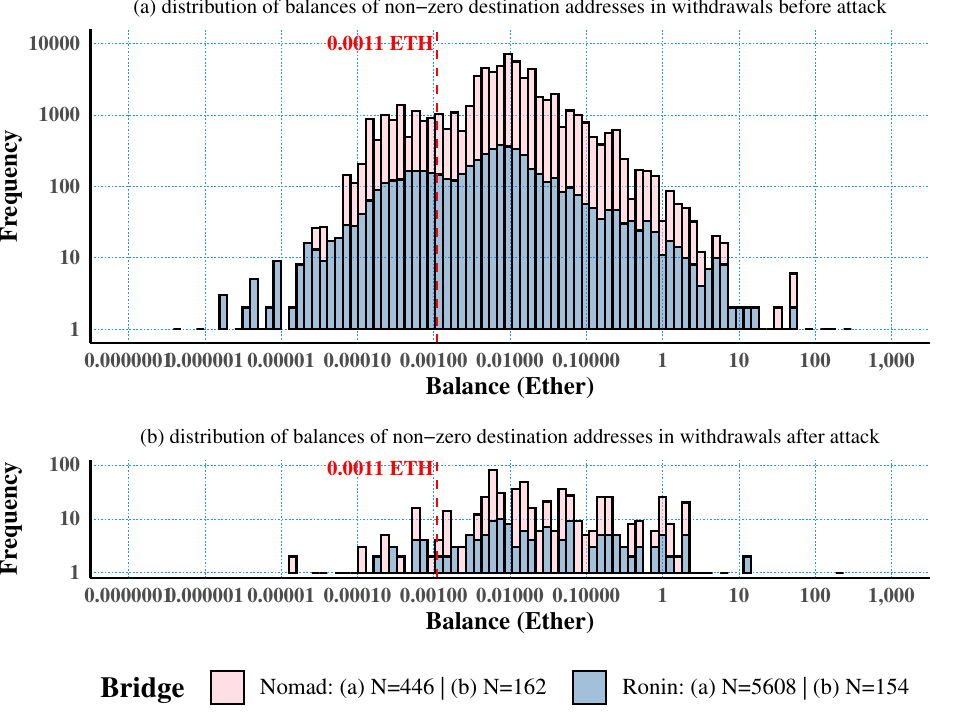}
    \caption{Distribution of the balance of all addresses to which funds are being sent in $\mathcal{S}$ when withdrawing funds from $\mathcal{T}$. A red dotted line marks the value needed to pay for transaction fees to successfully withdraw funds in $\mathcal{S}$.}
    \label{figure: destination-addresses-metrics-merged}
    \Description[<short description>]{<long description>}
\end{figure}

A manual analysis of these anomalies revealed that many of the destination addresses (beneficiaries on Ethereum) targeted by these events on $\mathcal{T}$ had no funds or had not made any transactions to date. Table~\ref{table: destination-addresses-metrics-merged} illustrates these findings, separating metrics extracted before and after (i.e., as a consequence of) the attack. Our analysis revealed that, spanning both bridges, 6,175 addresses on Ethereum ($\approx49\%$) had a zero balance at the time of the withdrawal event, in which 5,333 ($\approx43\%$) are still holding a zero balance at the time of writing. As a result, users cannot withdraw their assets due to not having funds to cover gas fees. 
According to the Ronin documentation, users should have a minimum of $0.0011$ ETH to cover gas fees for issuing a transaction on Ethereum to withdraw funds~\cite{ronin_fee}. 7,700 addresses ($\approx$ $61$) did not have sufficient funds to meet this requirement. The total value of unwithdrawn funds amounts to $\$4.8$M, in which a single transaction attempted to withdraw $\$3$M. Excluding this outlier, the amount not withdrawn is $\$1.8$M. Figure~\ref{figure: destination-addresses-metrics-merged} shows the distribution of balances of \textit{beneficiary} addresses with non-zero balances when the withdrawal event was triggered in $\mathcal{T}$. 

To assess the impact of the attack on these values, we divided the analysis into \emph{pre-attack} and \emph{post-attack} periods. The data shows that the attack does not seem to have any influence. The number of data points before the attack is much higher ($\approx$97\%), suggesting that this is a common practice when the bridge is operating normally. Interestingly, even users with many funds, including those with over $10$ or $200$ ETH in their addresses, were involved in this behavior (cf. Figure~\ref{figure: destination-addresses-metrics-merged}). 
Another noteworthy finding is the difference between unique addresses that attempted to withdraw funds once versus those that tried multiple times. Some users repeatedly attempted withdrawals, while others seemingly gave up, likely considering their funds lost. This may also be attributed to user inexperience and inadequate UI/UX~\cite{belchior_cacm}. The Pearson correlation coefficient between the number of withdrawal attempts and the amount withdrawn (by each user withdrawal request) is negligible ($-0.017$) showing no meaningful relationship between both variables.

\begin{anomalybox}
\textbf{Finding 7.} We found 729 cases where users tried to withdraw funds from the destination blockchain ($\mathcal{T}$), but the bridge never completed the corresponding transaction on the source blockchain ($\mathcal{S}$). This left up to \$4.8M stuck in the bridge. While many of these happened around the time of the attack, the majority occurred during normal use. Moreover, nearly half of the users didn’t have enough ETH to pay gas fees on the destination blockchain, preventing them from claiming their funds, and pointing to serious usability issues.
\end{anomalybox}

\subsubsection{Withdrawals in \texorpdfstring{$\mathcal{S}$}{S} with no Correspondence in \texorpdfstring{$\mathcal{T}$}{T}}
\label{subsec: withdrawals_in_s}

Both bridges analyzed in this paper suffered a \textbf{Forged Withdrawal Attack}, where funds were stolen from \(\mathcal{S}\) (Ethereum). As shown in Table~\ref{table: cctx_analysis}, $382$ unmatched events, under \RuleName{SC\_ValidERC20TokenWithdrawal}, were identified because they were not matched on \(\mathcal{T}\) on the Nomad bridge. 
Analyzing the timestamp of the transactions in which these events were emitted, we conclude that all 382 events were part of the attack, involving \textbf{$382$ transactions} and \textbf{$279$ unique addresses}. These totaled \textbf{$\$159,577,598$} of stolen funds. These events had only $14$ unique withdrawal IDs, indicating that attackers copy-pasted data from other transactions, exploiting the bridge's acceptance of any data as valid proof~\cite{nomad_rekt_2022}. Our analysis identified $279$ addresses that exploited the protocol, the majority of which were contracts deployed in bulk to scatter funds across multiple addresses. We traced the transactions and identified \textbf{45 unique EOAs} responsible for deploying these contracts. We cross-referenced our findings with data from \emph{Peckshield}, a reputable security firm, which provided a list of addresses involved in exploiting the bridge at the time of the attack~\cite{peckshield_dataset}. We identified $9$ more EOAs than Peckshield in the same blockchains ($36$ EOAs). We also found a dataset related to the attack on GitHub~\cite{github_nomad_xyz}, which includes $246$ transactions, less $136$ than ours. To eliminate the possibility of false positives, we manually checked all anomalies not identified by the other datasets.


In the Ronin bridge data, we identified 710 anomalies related to events emitted on $\mathcal{S}$ without correspondence on $\mathcal{T}$. Unfortunately, due to rate limits for extracting data from the Ronin blockchain, we could not decrease $t_0$ to the date on which the contracts were deployed. This caused our tool to identify anomalies in transactions that would match events emitted well before our period of analysis. We captured over 500k additional transactions in $[t_0; t_1[$, more than 3.5 months taking into account the maximum latency of withdrawals in the Ronin bridge (cf. Figure~\ref{figure: cctx_latency_vs_value_nomad}), of around 3 months, and added some margin, but it did not prove to be enough.
To exclude withdrawals before $t_0$, we based ourselves on the \Keyword{withdrawal\_id} -- a counter incremented for each withdrawal event emitted in the bridge contract. Of the unmatched $710$ events, $708$ had a withdrawal ID less than \Keyword{withdrawal\_id} of the first event included in $[t_0;t_1[$, suggesting that they were emitted before our collection data interval. We are left with $2$ unmatched withdrawal events in the selected interval. These events were emitted by transactions issued by the same address (\AddrHref{https://etherscan.io/address/0x098b716b8aaf21512996dc57eb0615e2383e2f96}{0x098b716b8aaf21512996dc57eb0615e2383e2f96}) on March 23, 2022, 13:29 (\AddrHref{https://etherscan.io/tx/0xc28fad5e8d5e0ce6a2eaf67b6687be5d58113e16be590824d6cfa1a94467d0b7}{0xc28fad5e8d5e0ce6a2eaf67b6687be5d58113e16be590824d6cfa1a94467d0b7}) and 13:31 (\AddrHref{https://etherscan.io/tx/0xed2c72ef1a552ddaec6dd1f5cddf0b59a8f37f82bdda5257d9c7c37db7bb9b08}{0xed2c72ef1a552ddaec6dd1f5cddf0b59a8f37f82bdda5257d9c7c37db7bb9b08}), transferring a total of $\$565.64$M. These two transactions are those identified in the industry as pertaining to the Ronin bridge hack. When comparing the results, no false negatives were found.

\begin{anomalybox}
\textbf{Finding 8.} \toolName~successfully identified malicious transactions in the Nomad and Ronin bridges. The Nomad analysis overcame previous analyses by uncovering 9 additional attacker EOAs not reported by security firms and 136 more transactions than the largest existing public dataset of the Nomad hack.
\end{anomalybox}

%% file: tables/time-intervals.tex
\setlength{\tabcolsep}{10pt}

\begin{table}[ht]
\centering
\scriptsize
\caption{Timeframes of Relevance for Data Extraction}
\begin{tabular}{@{}lcccc@{}}
\toprule
 & {\small $t_0$} & {\footnotesize $t_1$} & {\footnotesize $t_2$} & {\footnotesize $t_3$} \\ \midrule
Nomad Bridge & -- & \begin{tabular}[c]{@{}c@{}}Jan 11, 2022\\ (1641905876)\end{tabular} & \begin{tabular}[c]{@{}c@{}}Dec 15, 2022\\ (1671062400)\end{tabular} & \begin{tabular}[c]{@{}c@{}}Jul 31, 2024\\ (1722441775)\end{tabular} \\
\addlinespace[0.1cm]
Ronin Bridge & \begin{tabular}[c]{@{}c@{}}Sep 13, 2021\\ (1631491200)\end{tabular} & \begin{tabular}[c]{@{}c@{}}Jan 1, 2022\\ (1640995200)\end{tabular} & \begin{tabular}[c]{@{}c@{}}Apr 28, 2022\\ (1651156446)\end{tabular} & \begin{tabular}[c]{@{}c@{}}Jul 31, 2024\\ (1722441775)\end{tabular} \\ \bottomrule
\end{tabular}\label{table: timestamps}
\begin{tablenotes}
 \footnotesize
 \item \textit{Note:} The interval of interest is $[t_1,t_2]$. 
 The table presents dates and corresponding Unix timestamps in parentheses. The Nomad and Ronin bridges were attacked on Aug 2, 2022 and Mar 22, 2022, respectively.
\end{tablenotes}
\end{table}

%% file: tables/evaluation.tex
\setlength{\tabcolsep}{5pt}

\begin{table}[t]
\centering
\footnotesize
\caption{Facts extraction latency (in seconds) per token type}
\begin{tabular}{@{}cccccccc@{}}
\toprule
Bridge & Token type & size & min & max & avg & median & std \\
\midrule
\multirow{2}{*}{Ronin} & native     & $468,997$ & $0.18$ & $138.15$ & $1.82$ & $0.35$ & $4.70$ \\

 & non-native & $347,580$ & $3.81$x$10^{-6}$ & $3.65$  & $0.28$ & $0.23$ & $0.26$ \\

\arrayrulecolor{lightgray}

\hdashline
\addlinespace[0.1cm]

\multirow{2}{*}{Nomad} & native & $7,656$ & $0.16$ & $8.78$ & $0.89$ & $0.78$  & $0.46$ \\
 & non-native & $51,702$ & $3.81$x$10^{-6}$ & $5.83$ & $0.26$ & $0.19$ & $0.28$ \\

\arrayrulecolor{black}

\bottomrule
\end{tabular}\label{table: evaluation}
\end{table}

%% file: tables/anomaly-detection-analysis.tex
\setlength{\tabcolsep}{2pt} 

\begin{table*}[ht]

\centering
\footnotesize
\caption{Anomaly detection results, within $[t_1;t_2]$, using the cross-chain rules defined in Section~\ref{subsec: cross_chain_rules}}
\begin{threeparttable}[c]
\begin{tabular}{lcccc}
\toprule
 & \multicolumn{2}{c}{Nomad Bridge} & \multicolumn{2}{c}{Ronin Bridge} \\
\cmidrule(rl){2-3} \cmidrule(rl){4-5} 


Logical Rule (cf. Section~\ref{subsec: cross_chain_rules}) & Captured Records & Anomalies Detected & Captured Records & Anomalies Detected \\ \midrule

\addlinespace[0.05cm]

\rowcolor{gray!10}1. \ScriptsizeKeyword{SC\_ValidNativeTokenDeposit}  
& 7,187  
& 0
& 38,462  
& 0 \\

\rowcolor{white}2. \ScriptsizeKeyword{SC\_ValidERC20TokenDeposit}
& 4,223 
& 39 (14 phishing attempts + 25 transfers to bridge)
& {5527}
& 83 (3 phishing attempts + 80 transfers to bridge) \\
\multicolumn{2}{c}{\textbf{Total Value in Transfers to Bridge }} & \textbf{\$93.86K} & & \textbf{\$113.00K} \\

\rowcolor{gray!10}3. \ScriptsizeKeyword{TC\_ValidERC20TokenDeposit} 
& 11,417 
& 0
& 43,990  
& 0   \\

\rowcolor{white}4. \ScriptsizeKeyword{CCTX\_ValidDeposit} 
& 11,404 
& 19\tnote{*} 
& 43,979  
& 10\tnote{*} \\

\addlinespace[0.05cm]
\arrayrulecolor{lightgray}
\hline
\addlinespace[0.05cm]


\rowcolor{gray!10}5. \ScriptsizeKeyword{TC\_ValidNativeTokenWithdrawal}
& 464
& 0
& 0
& 0 \\

\rowcolor{white}6. \ScriptsizeKeyword{TC\_ValidERC20TokenWithdrawal}
& 4,846
& 10 (3 unparseable addresses + 7 attack attempts)
& 35,413
& 0 \\

\rowcolor{gray!10}7. \ScriptsizeKeyword{SC\_ValidERC20TokenWithdrawal}
& 4,869
& 2 (2 phishing attempts)
& 25,470
& 1 (1 phishing attempt) \\

\rowcolor{white}8. \ScriptsizeKeyword{CCTX\_ValidWithdrawal}
& 4,482
& 729\tnote{*}
& 22,830
& 12,546\tnote{*} \\

\arrayrulecolor{black}
\bottomrule
\end{tabular}
\begin{tablenotes}
    \item Recall that the rules capture expected behavior. Therefore, the anomalies presented are the result of comparing each event emitted by each contract, with being captured or not by the corresponding rule that should have captured it.
   \item [*] Table~\ref{table: cctx_analysis} presents a detailed explanation of these anomalies. Each anomaly is categorized based on the underlying reasons that led to its occurrence.
\end{tablenotes}
\end{threeparttable}
\label{table: anomaly_detection_results}
\end{table*}

%% file: tables/deposits_and_withdrawals_anomalies.tex
\setlength{\tabcolsep}{5pt} 

\begin{table*}[ht]

\centering
\scriptsize
\caption{Identification of the origin of all anomalies identified by \FootNoteSizeRuleName{CCTX\_ValidDeposit} and \FootNoteSizeRuleName{CCTX\_ValidWithdrawal} within $[t_0;t_3]$}
\begin{threeparttable}[c]

\begin{tabular}{lcclccl}

\toprule
 & \multicolumn{3}{c}{Nomad Bridge} & \multicolumn{3}{c}{Ronin Bridge} \\
\cmidrule(rl){2-4} \cmidrule(rl){5-7} 

Logical Rule (cf. Section~\ref{subsec: cross_chain_rules}) & Captured & Unmatched & Anomaly Explanation & Captured & Unmatched & Anomaly Explanation \\ \midrule

1. \ScriptsizeKeyword{SC\_ValidNativeTokenDeposit}  
& \SCValidNativeTokenDepositNomadCaptured  
& \SCValidNativeTokenDepositNomadUnmatched
& 
& \SCValidNativeTokenDepositRoninCaptured  
& \SCValidNativeTokenDepositRoninUnmatched
& 10 do not comply with \scriptSizeRuleName{cctx\_finality} \\

\addlinespace[0.05cm]
\arrayrulecolor{lightgray}
\hdashline
\addlinespace[0.05cm]

\multirow{2}{*}{2. \ScriptsizeKeyword{SC\_ValidERC20TokenDeposit}}
& \multirow{2}{*}{\SCValidERCTokenDepositNomadCaptured}
& \multirow{2}{*}{\SCValidERCTokenDepositNomadUnmatched}
& 5 do not comply with \scriptSizeRuleName{cctx\_finality} 
& \multirow{2}{*}{\SCValidERCTokenDepositRoninCaptured}
& \multirow{2}{*}{\SCValidERCTokenDepositRoninUnmatched}
&  \\

& 
& 
& 1 contains an invalid beneficiary address (FP)
& 
& 
& \\

\addlinespace[0.05cm]
\arrayrulecolor{lightgray}
\hdashline
\addlinespace[0.05cm]

\multirow{3}{*}{3. \ScriptsizeKeyword{TC\_ValidERC20TokenDeposit} }
& \multirow{3}{*}{\TCValidERCTokenDepositNomadCaptured }
& \multirow{3}{*}{\TCValidERCTokenDepositNomadUnmatched}
& 5 do not comply with \scriptSizeRuleName{cctx\_finality} 
& \multirow{3}{*}{\TCValidERCTokenDepositRoninCaptured  }
& \multirow{3}{*}{10}
& \multirow{3}{*}{10 do not comply with \scriptSizeRuleName{cctx\_finality}}  \\

& 
& 
& 7 do not comply with \scriptSizeRuleName{token\_mapping}
& 
& 
& \\

& 
& 
& 1 contains an invalid beneficiary address (FP)
& 
& 
& \\

\addlinespace[0.05cm]
\arrayrulecolor{lightgray}
\hdashline
\addlinespace[0.05cm]

5. \ScriptsizeKeyword{TC\_ValidNativeTokenWithdrawal}
& \TCValidNativeTokenWithdrawalNomadCaptured
& \TCValidNativeTokenWithdrawalAddDataNomadUnmatched
& 238 events do not have correspondence on \(\mathcal{S}\)
& \TCValidNativeTokenWithdrawalRoninCaptured
& \TCValidNativeTokenWithdrawalRoninUnmatched
&  \\

\addlinespace[0.05cm]
\arrayrulecolor{lightgray}
\hdashline
\addlinespace[0.05cm]

\multirow{2}{*}{6. \ScriptsizeKeyword{TC\_ValidERC20TokenWithdrawal}}
& \multirow{2}{*}{\TCValidERCTokenWithdrawalNomadCaptured}
& \multirow{2}{*}{\TCValidERCTokenWithdrawalAddDataNomadUnmatched}
& \multirow{2}{*}{491 events do not have correspondence on \(\mathcal{S}\)}
& \multirow{2}{*}{\TCValidERCTokenWithdrawalRoninCaptured}
& \multirow{2}{*}{11,814}
& 22 do not comply with \scriptSizeRuleName{cctx\_finality} \\

& 
& 
& 
& 
& 
& 11,792 events do not have correspondence on \(\mathcal{S}\) \\

\addlinespace[0.05cm]
\arrayrulecolor{lightgray}
\hdashline
\addlinespace[0.05cm]

\multirow{3}{*}{7. \ScriptsizeKeyword{SC\_ValidERC20TokenWithdrawal}}
& \multirow{3}{*}{\SCValidERCTokenWithdrawalNomadCaptured}
& \multirow{3}{*}{\SCValidERCTokenWithdrawalNomadUnmatched}
& 3 contains an invalid beneficiary address (FP)
& \multirow{3}{*}{\SCValidERCTokenWithdrawalRoninCaptured}
& \multirow{3}{*}{732}
& 708 matched events on \(\mathcal{T}\) before $t_0$ (FP)\tnote{1}\\

& 
& 
& 2 do not comply with \scriptSizeRuleName{token\_mapping}
& 
& 
& 22 do not comply with \scriptSizeRuleName{cctx\_finality}  \\

& 
& 
& \cellcolor{red!50}382 events do not have correspondence on \(\mathcal{T}\)
& 
& 
& \cellcolor{red!50}2 events do not have correspondence on \(\mathcal{T}\) \\

\bottomrule
\end{tabular}

\begin{tablenotes}
 \footnotesize
 \item \textit{Example:} there were \TCValidERCTokenDepositNomadCaptured records captured by \FootNoteSizeRuleName{TC\_ValidERC20TokenDeposit} (Rule 3), however, only \CCTXDepositNomadCaptured were matched by a transaction on $\mathcal{S}$ (counted in \FootNoteSizeRuleName{CCTX\_ValidDeposit} -- cf. Table~\ref{table: anomaly_detection_results}), which indicates there are 13 events emitted by the bridge contract on $\mathcal{T}$ without a corresponding action on $\mathcal{S}$, which is an anomaly.
 \item \textit{Note:} we mark in red the events that caused loss of funds to the protocol (i.e., attacks identified by \toolName)
 \item [1] false positives (FP) due to the impossibility of extracting data in the Ronin blockchain ($\mathcal{T}$) before $t_0$, which caused the events to not being matched
\end{tablenotes}
\end{threeparttable}

\label{table: cctx_analysis}
\end{table*}

%% file: tables/destination-addresses-metrics.tex
\setlength{\tabcolsep}{10pt} 

\begin{table*}[ht]
\rowcolors{2}{gray!10}{white}
\footnotesize
\caption{Analysis of the balance of destination addresses on Ethereum targeted by withdrawals on $\mathcal{T}$}
\begin{threeparttable}[c]
\begin{tabular}{lcccccc}
\toprule
 & \multicolumn{3}{c}{Nomad Bridge} & \multicolumn{3}{c}{Ronin Bridge} \\
 \cmidrule(rl){2-4} \cmidrule(rl){5-7} 

 & Before Attack & After Attack & \textbf{Total} & Before Attack & After Attack & \textbf{Total} \\
\midrule
Unmatched withdrawal events in $\mathcal{T}$ &
\totaldstaddressesinwithdrawalswithnomatchbeforeNomad &
\totaldstaddressesinwithdrawalswithnomatchafterNomad &
\textbf{\totaldstaddressesinwithdrawalswithnomatchNomad} &
\totaldstaddressesinwithdrawalswithnomatchbeforeRonin &
\totaldstaddressesinwithdrawalswithnomatchafterRonin &
\textbf{\totaldstaddressesinwithdrawalswithnomatchRonin} \\

Addresses with balance $0$ at withdrawal date &
\totaldstaddressesinwithdrawalswithnomatchbeforenobalanceNomad &
\totaldstaddressesinwithdrawalswithnomatchafternobalanceNomad &
\textbf{\totaldstaddressesinwithdrawalswithnomatchnobalanceNomad} &
\totaldstaddressesinwithdrawalswithnomatchbeforenobalanceRonin &
\totaldstaddressesinwithdrawalswithnomatchafternobalanceRonin &
\textbf{\totaldstaddressesinwithdrawalswithnomatchnobalanceRonin} \\

Addresses with balance $0$ at withdrawal date and still today &
\totaldstaddressesinwithdrawalswithnomatchbeforenobalanceandunderfeeNomad &
\totaldstaddressesinwithdrawalswithnomatchafternobalanceandunderfeeNomad &
\textbf{\totaldstaddressesinwithdrawalswithnomatchnobalanceandunderfeeNomad} &
\totaldstaddressesinwithdrawalswithnomatchbeforenobalanceandunderfeeRonin &
\totaldstaddressesinwithdrawalswithnomatchafternobalanceandunderfeeRonin &
\textbf{\totaldstaddressesinwithdrawalswithnomatchnobalanceandunderfeeRonin} \\

Addresses with balance $<\;0.0011$ at withdrawal date &
\totaldstaddressesinwithdrawalswithnomatchbeforeunderfeeNomad&
\totaldstaddressesinwithdrawalswithnomatchafterunderfeeNomad &
\textbf{\totaldstaddressesinwithdrawalswithnomatchunderfeeNomad} &
\totaldstaddressesinwithdrawalswithnomatchbeforeunderfeeRonin &
\totaldstaddressesinwithdrawalswithnomatchafterunderfeeRonin &
\textbf{\totaldstaddressesinwithdrawalswithnomatchunderfeeRonin} \\
Total Value (in million of USD) &
\$\totaldstaddressesinwithdrawalswithnomatchbeforetotalvalueNomad &
\$\totaldstaddressesinwithdrawalswithnomatchaftertotalvalueNomad\tnote{1} &
\textbf{\$\totaldstaddressesinwithdrawalswithnomatchtotalvalueNomad}\tnote{1} &
\$\totaldstaddressesinwithdrawalswithnomatchbeforetotalvalueRonin &
\$\totaldstaddressesinwithdrawalswithnomatchaftertotalvalueRonin &
\textbf{\$\totaldstaddressesinwithdrawalswithnomatchtotalvalueRonin} \\
 
Addresses that tried withdrawing more than once &
\totaldstaddressesinwithdrawalswithnomatchbeforemultipletriesNomad &
\totaldstaddressesinwithdrawalswithnomatchaftermultipletriesNomad &
\textbf{\totaldstaddressesinwithdrawalswithnomatchmultipletriesNomad} &
\totaldstaddressesinwithdrawalswithnomatchbeforemultipletriesRonin &
\totaldstaddressesinwithdrawalswithnomatchaftermultipletriesRonin &
\textbf{\totaldstaddressesinwithdrawalswithnomatchmultipletriesRonin} \\
 
Addresses that tried withdrawing exactly once &
\totaldstaddressesinwithdrawalswithnomatchbeforesingletriesNomad &
\totaldstaddressesinwithdrawalswithnomatchaftersingletriesNomad &
\textbf{\totaldstaddressesinwithdrawalswithnomatchsingletriesNomad} &
\totaldstaddressesinwithdrawalswithnomatchbeforesingletriesRonin &
\totaldstaddressesinwithdrawalswithnomatchaftersingletriesRonin &
\textbf{\totaldstaddressesinwithdrawalswithnomatchsingletriesRonin} \\
 
\bottomrule
\end{tabular}
\begin{tablenotes}
   \item [1] A single address is responsible for \$3M.
\end{tablenotes}
\end{threeparttable}
\label{table: destination-addresses-metrics-merged}
\end{table*}

%% file: sections/limitations-and-future-work.tex
\section{Discussion and Future Work}
\label{sec: limitations_and_future_work}

We present the discussion and limitations of our work.

\featheader{Rule modeling} Modeling cross-chain rules requires exploring the semantics of a bridge protocol, its data model, and associated token contracts. While bridges can be categorized into several classes of models -- in this paper, we analyzed bridges that connect Ethereum to sidechains. Specific rule modeling may change slightly depending on the particular instantiation. This paper does not aim to propose a universal cross-chain model applicable to all protocols. Instead, we empirically demonstrate that logic-driven analysis is effective for the detection of unknown anomalies on bridges. Our work establishes the first baseline for future security analyses on cross-chain bridges. 
Rules are created based on the current behavior of the protocol. If event signatures are changed, \toolName~needs to be updated accordingly by the bridge operator to capture the new events. This seems acceptable as \toolName~is supposed to run alongside the bridge and controlled by the operator.


\featheader{Timeframes and Selected Bridges} We focus on short timeframes with confirmed attacks, modeling expected behavior rather than relying on signature-based detection. Since there are no prior anomaly datasets for cross-chain bridges, manually analyzing large volumes of anomalies would be impractical. Thus, we target (1) short periods and (2) timeframes with verified attacks. We selected the Ronin bridge as it is the most profitable cross-chain attack to date, and the Nomad bridge because of the high number of transactions exploiting the protocol in the hack.

\featheader{Extensibility of \toolName} 
The framework is extensible and easy to use. To add support for other protocols, users must \textbf{(1)} analyze the protocol and incorporate any specific restrictions into the cross-chain rules (i.e., adding any missing protocol-specific cross-chain rules), \textbf{(2)} extract transaction receipts to be used in the analysis, \textbf{(3)} create an \textit{Event Data Decoder and Extractor} that decodes event data and creates logical relations, and \textbf{(4)} populate a configuration file (cf. Figure~\ref{fig: cross-chain-rules-validator-pipeline}) with RCP connection URLs, bridge contract addresses, and tokens mappings. An important feature of the proposed design of \toolName~is that it is agnostic to the state validation logic employed -- \emph{Trusted Third Parties} (Ronin), and \emph{Native State Verification} (Nomad)~\cite{augusto_sok_2024} -- because it only relies on the events emitted by on-chain contracts.


\featheader{Event-based Analysis}
We chose to perform an event-based analysis for two main reasons. Firstly, protocols typically involve more transactions than those triggering transferring assets, e.g., light client updates. Analyzing all transactions, including those unrelated to actual state changes, would be inefficient and resource-intensive. 
Moreover, capturing all transactions that target bridge contracts is not enough to extract all the relevant data, as users can issue transactions to intermediary protocols (such as bridge aggregators~\cite{Subramanian_2024}) that make internal transaction calls to bridge contracts. 
When contract events are not emitted (e.g., due to a bug in a contract or even a malicious upgrade in a \textbf{multi-transaction attack}), our tool detects this behavior as abnormal because a state change will be missing in the cross-chain flow (cf. Section~\ref{sec: cross-chain-bridge-model}). 

\featheader{Future Work}
Future work is threefold: \textbf{(1)} extend analysis periods to identify further anomalies, such as \emph{salami slicing attacks}~\cite{bhowmik2008data}, \textbf{(2)} support additional bridges, \textbf{(3)} using the clean and labeled dataset to train anomaly detection models to perform large scale analyses of cross-chain data.

%% file: sections/related-work.tex
\section{Related Work}
\label{sec: related_work}

Despite the extensive research corpus on interoperability~\cite{belchior2021survey, belchior_cacm, Zamyatin_xclaim_2019}, there is little related work available on monitoring and protecting interoperability solutions. The concept of a cross-chain model was introduced in Hephaestus~\cite{belchior_hephaestus_2022}, a theoretical cross-chain model generator, highlighting the importance of defining cross-chain rules to identify misbehavior. 
XScope~\cite{zhang_xscope_2022} uses three static rules to detect three types of attacks (signature-based detection) on cross-chain bridges, specifically targeting smaller chains with limited datasets. 
XScope's detection capabilities are limited to three specific anomalies, and it focuses exclusively on token deposits (not covering the withdrawal process). Unfortunately, XScope is not open-source, limiting a deeper empirical comparison.
In the industry, Hyperlane~\cite{hyperlane}, Range~\cite{range}, and Layer Zero's Precrime~\cite{precrime} provide analysis tools for bridges. However, these are proprietary and lack technical documentation, systematic evaluation, and datasets, making it challenging to compare directly with our work. 
Finally, while post-attack analyses typically trace the flow of funds using tools such as Chainalysis~\cite{chainalysis}, our tool enables the retrieval of the same (and more) data by applying cross-chain rules. 

%% file: sections/conclusion.tex
\section{Conclusion}
\label{sec: conclusion}

This paper proposes a monitoring framework for cross-chain bridges powered by a cross-chain model supported by a Datalog engine. We \emph{uncover significant attacks} within cross-chain bridges, such as 1) transactions accepted in one chain \emph{before the finality time of the original one elapsed}, breaking the safety of the bridge protocol; 2) users trying to exploit a protocol through \emph{the creation of fake versions of wrapped Ether} to withdraw real ether on the Ethereum blockchain, breaking safety; 3) bridge contract implementations handling \emph{unexpected inputs differently across chains}, hindering a good UX and leading to the loss of user funds. In addition, we identify every transaction involved in previous hacks on the bridges studied. 
We show that although only $49$ unique externally owned accounts (EOAs) exploited Nomad, there were $380$ exploit events, with each address deploying multiple exploit contracts to obscure the flow of funds. Finally, our study highlights a critical user awareness gap -- many users struggle to withdraw funds due to the highly manual nature of the process, contrasting with the more streamlined deposit process managed by bridge operators. This user error has led to over $\$4.8$M in unwithdrawn funds due to users mistakenly sending funds to addresses they do not control or that have never been active. We are the first to empirically analyze the security vulnerabilities of cross-chain bridges. Our open-source dataset provides a valuable resource for future research.